\documentclass[conference]{IEEEtran}
\usepackage{cite}
\usepackage{amsmath}
\usepackage{amssymb,amsfonts}
\usepackage{graphicx}
\usepackage{textcomp}
\usepackage{xcolor}
\usepackage{tikz}
\usepackage[hyphens]{url}
\usepackage{fancyhdr}
\usepackage[bookmarks=true,breaklinks=true,letterpaper=true,colorlinks,citecolor=blue,linkcolor=blue,urlcolor=blue]{hyperref}
\usepackage{makecell}
\usepackage[ruled,linesnumbered,vlined]{algorithm2e}
\usepackage{balance}

\newcommand{\ourdesign}{STAR}

\newcommand*\circled[1]{\tikz[baseline=(char.base)]{
    \node[shape=circle,draw,fill=black,text=white,font=\bf,inner sep=0.5pt] (char)
  {\scriptsize#1};
}}

\newif\ifremark
\long\def\remark#1{
\ifremark%
        \begingroup%
        \dimen0=\columnwidth
        \advance\dimen0 by -1in%
        \setbox0=\hbox{\parbox[b]{\dimen0}{\protect\em #1}}
        \dimen1=\ht0\advance\dimen1 by 2pt%
        \dimen2=\dp0\advance\dimen2 by 2pt%
        \vskip 0.25pt%
        \hbox to \columnwidth{%
                \vrule height\dimen1 width 3pt depth\dimen2%
                \hss\copy0\hss%
                \vrule height\dimen1 width 3pt depth\dimen2%
        }%
        \endgroup%
\fi}

\remarktrue

\title{Improving Multi-Instance GPU Efficiency via Sub-Entry Sharing TLB Design} 

\author{\IEEEauthorblockN{Bingyao Li\IEEEauthorrefmark{2},
Yueqi Wang\IEEEauthorrefmark{2}, Tianyu Wang\IEEEauthorrefmark{2}, Lieven Eeckhout\IEEEauthorrefmark{3}, Jun Yang\IEEEauthorrefmark{2}, Aamer Jaleel\IEEEauthorrefmark{4}, Xulong Tang\IEEEauthorrefmark{2}}
\IEEEauthorblockA{University of Pittsburgh\IEEEauthorrefmark{2}, Ghent University\IEEEauthorrefmark{3}, NVIDIA\IEEEauthorrefmark{4} }
\IEEEauthorblockA{Email: \IEEEauthorrefmark{2}\{bil35, yuw249, tiw81, juy9, tax6\}@pitt.edu,
\IEEEauthorrefmark{3}Lieven.Eeckhout@ugent.be,
\IEEEauthorrefmark{4}ajaleel@nvidia.com
}

}

\begin{document}
\maketitle
\pagestyle{plain}

\newcommand{\squishlist}{
    \begin{list}{$\bullet$}
        { \setlength{\itemsep}{0pt}      \setlength{\parsep}{0pt}
            \setlength{\topsep}{0.5pt}       \setlength{\partopsep}{0pt}
            \setlength{\listparindent}{-2pt}
            \setlength{\itemindent}{-5pt}
            \setlength{\leftmargin}{0.5em} \setlength{\labelwidth}{0em}
            \setlength{\labelsep}{0.2em} } }
    
\newcommand{\squishend}{
\end{list}  }


\begin{abstract}

NVIDIA's Multi-Instance GPU (MIG) technology enables partitioning GPU computing power and memory into separate hardware instances, providing complete isolation including compute resources, caches, and memory. However, prior work identifies that MIG does not extend to partitioning the last-level TLB (i.e., L3 TLB), which remains shared among all instances. To enhance TLB reach, NVIDIA GPUs reorganized the TLB structure with 16 sub-entries in each L3 TLB entry that have a one-to-one mapping to the address translations for 16 pages of size 64\,KB located within the same 1\,MB aligned range. 
Our comprehensive investigation of address translation efficiency in MIG identifies two main issues caused by L3 TLB sharing interference: (i) it results in performance degradation for co-running applications, and (ii) TLB sub-entries are not fully utilized before eviction. Based on this observation, we propose \ourdesign~to improve the utilization of TLB sub-entries through dynamic sharing of TLB entries across multiple base addresses. \ourdesign~evaluates TLB entries based on their sub-entry utilization to optimize address translation storage, dynamically adjusting between a shared and non-shared status to cater to current demand. We show that \ourdesign~improves overall performance by an average of 30.2\% across various multi-tenant workloads.

\end{abstract}

\section{Introduction}


Graphics Processing Units (GPUs) are extensively utilized in contemporary computing systems to accelerate performance across various applications. As artificial intelligence/ML models evolve, GPU manufacturers are continually enhancing the capabilities of individual GPUs to meet the surging computational demands~\cite{yi2020heimdall, zhao2021holoar,feng2019parallelism, mining, mining2, 10070956}. However, previous research has shown that these advanced applications still cannot fully exploit the existing GPU computational resources due to different workloads facing various resource bottlenecks and exhibiting different sensitivities to different resources~\cite{li2022miso, kim2022paris, yu2021automated, multi-tenancy-3, multi-tenancy-2, multi-tenancy-1}.

To address the issue of underutilization,  GPU vendors are evolving to offer GPU resource partitioning capabilities to enable multiple applications to share the same physical GPU resources. NVIDIA's Multi-Instance GPU (MIG)~\cite{nvidia-mig-docu} is one of the prominent GPU-sharing technologies. MIG enables a single physical GPU to be divided into several isolated instances, each with their own set of resources, including streaming multiprocessors (SMs), local memory, and caches. NVIDIA MIG is designed to offer isolation of resources for each instance, ensuring performance without interference from other instances. However, a recent study~\cite{zhang2023t} has indicated that while MIG effectively partitions most of the memory system, it does not partition the last-level TLB (i.e., L3 TLB).
The shared L3 TLB allows the TLB to dynamically allocate its entries based on the demand from various instances, optimizing the use of the available TLB capacity. On the flip side, TLB sharing also leads to contention among multi-tenant applications.

With the increasingly large data sets and wide memory footprints of applications, the TLB has become a critical performance bottleneck~\cite{6835964, talluri1994surpassing, 8192492, colt, li2021improving, HPCAli}. Expanding the TLB size to alleviate this issue is impractical due to hardware size constraints. In response, NVIDIA's new generation GPU (e.g., A100) presents an innovative TLB architecture to enhance TLB reach. Specifically, in the L2 and L3 TLBs, an entry comprises 16 sub-entries, each directly corresponding to the address translation of 16 sequential 64\,KB pages within a contiguous 1\,MB-aligned segment, as recently revealed through reverse-engineering~\cite{zhang2023t}.   
By compressing multiple translations into a single TLB entry, the TLB can manage more data with fewer entries, thereby reducing hardware overhead, while improving TLB efficiency and boosting overall performance. A sub-entry TLB design performs well for isolated workloads that use large contiguous memory, however, in multi-tenant setups where the L3 TLB is shared, this design can lead to sub-entry underutilization because interference from co-runners causes frequent evictions when only a portion of the sub-entries are used.



\begin{figure}[!t]
	\centering
\includegraphics[width=0.48\textwidth]{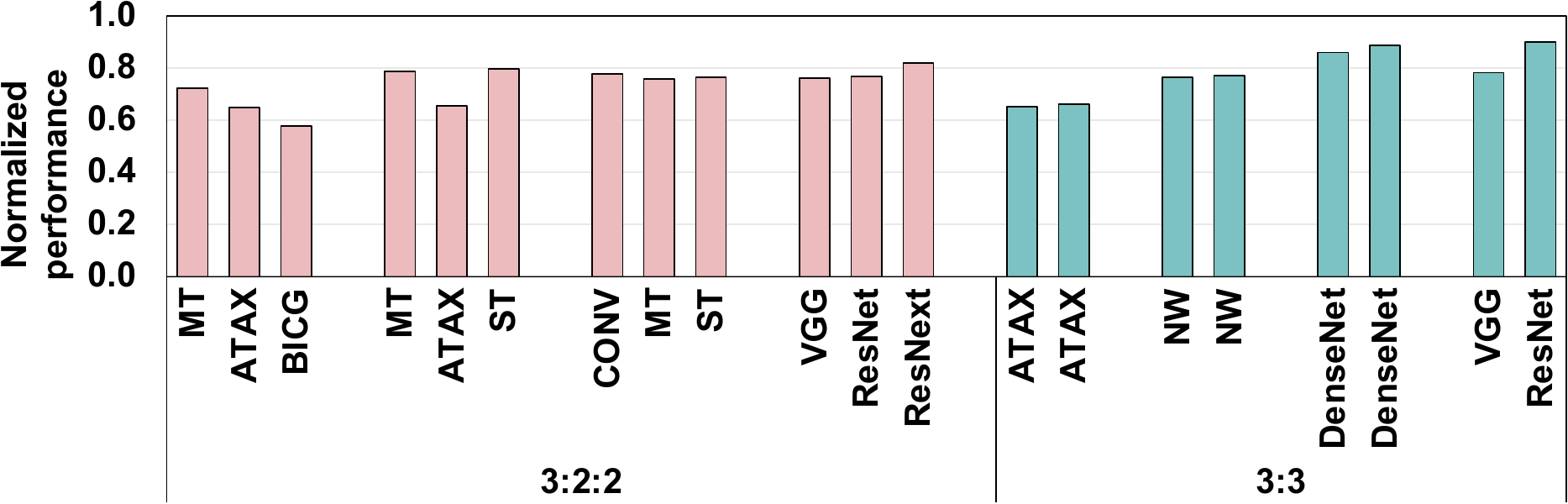}
\vspace{-8pt}
\caption{Performance of co-running applications on NVIDIA's A100. }

\label{fig:real_platform}
\vspace{-16pt}
\end{figure}

To understand the impact of L3 TLB contention in a multi-tenant environment, we co-run representative GPU applications on an NVIDIA A100 GPU with MIG enabled, see Figure~\ref{fig:real_platform} where each application runs within its own instance while sharing the L3 TLB. The GPU is partitioned into varying sizes of instances, including (3g, 2g, 2g) and (3g, 3g), where `g' represents the allocation of computing resources; each instance runs a single application. Performance is normalized to each application running alone on its respective instance, thereby having exclusive access to the L3 TLB. We observe that L3 TLB contention significantly degrades the performance of individual applications. This is because high access demand and interference from co-running applications lead to severe TLB thrashing. This thrashing extends the reuse distance of address translations, making translations less likely to be reused before they are evicted. It also leads to lower sub-entry utilization at the point of eviction, as the interference from concurrent requests accelerates TLB eviction (quantitative results and detailed analysis are given in Section~\ref{sec:motivation}).

A large body of prior work focused on improving address translation efficiency from multiple perspectives, including contiguity-based range TLBs~\cite{Karakostas:2015:RMM:2749469.2749471,8192492,Yan:2019:TRO:3307650.3322223, lee2023snakebyte},
cluster TLBs~\cite{6835964,colt}, employing large pages~\cite{8686553,8416828,pham2015large}, and TLB compression~\cite{10.1145/3410463.3414633}, TLB speculation~\cite{6307767}. Many of these optimizations are designed for single GPU/CPU setups running one application and are not effectively applicable to MIG environments with multiple tenants.
First, range-TLB, cluster-TLB, and TLB-compression strategies are optimized for sequential and stride memory access patterns, commonly found within individual applications. Co-running applications often have varied and unpredictable access patterns, making it challenging for these TLB optimizations to consistently capture the requested translations efficiently. Second, TLB speculation relies on the assumption of consistent access patterns to achieve accurate predictions. Similarly, in scenarios where the L3 TLB is accessed by multiple applications simultaneously, the interference between applications disrupts the regularity of memory accesses, significantly diminishing prediction accuracy. Third, using large pages can increase TLB reach by reducing the number of TLB entries needed to cover the same memory range. However, multi-tenant environments often host a mix of applications with diverse memory access patterns. While some applications can efficiently leverage large pages, others with irregular or sparse access patterns may not observe the same benefits. This variance leaves those less suited to large pages still facing contention issues. Other work, for example MASK~\cite{Ausavarungnirun:2018:MRG:3173162.3173169}, improves address translation efficiency in multi-application environments by controlling warp access to the shared TLB through an epoch- and token-based scheme. Although this approach is effective at reducing TLB thrashing, it helps little with TLB sub-entry utilization.

Motivated by these challenges, we systematically investigate and optimize the address translation in MIG systems. Our quantitative analysis reveals that contention in the L3 TLB critically undermines MIG performance, primarily due to low utilization of TLB sub-entries caused by multi-tenant interference. To address this, we propose the \textbf{S}ub-En\textbf{T}ry Sh\textbf{A}ring-Awa\textbf{R}e (\ourdesign) TLB, which dynamically allows different base addresses to share TLB entries. Specifically, instead of defaulting to Least Recently Used (LRU) eviction upon receiving a new address translation, our method evaluates and selects an entry based on its current sub-entry utilization that satisfies the sharing criteria for inserting the new address translation. Additionally, our approach can dynamically switch between a TLB entry's shared and non-shared states, adapting to the fluctuating demands of TLB sub-entries. We make the following contributions:

\squishlist{}

\item We show that a major performance bottleneck in MIG arises from severe contention in the shared L3 TLB. We provide a detailed analysis of how multi-tenant interference affects address translation reuse and TLB sub-entry utilization. 

\item We propose \ourdesign, a hardware design tailored to mitigate the negative effects of multi-tenant interference and enhance overall application performance. \ourdesign~enables multiple base addresses to share the same TLB entry, enhancing sub-entry utilization. It also dynamically switches between shared and non-shared states to adapt to varying application demands.

\item We show that \ourdesign~improves overall performance by an average of 30.2\%  across a suite of multi-tenant workloads. We show that \ourdesign~outperforms various TLB design alternatives and is orthogonal to these approaches to achieve further performance improvement.

\squishend{}
\section{Background}
\subsection{ Multi-Instance GPU}
\label{subsec:mig}
Modern GPUs, such as NVIDIA's Ampere and Hopper generations (e.g., A100 and H100), leverage Multi-Instance GPU (MIG) technology to enhance resource utilization by enabling the sharing and partitioning of GPU resources~\cite{A100, H100}.
MIG technology allows a single GPU to be divided into multiple GPU partitions, each operating as an independent GPU instance with its own dedicated resources. 
The partitioning includes SMs and the entire memory system, including the on-chip crossbar ports, L2 cache banks, memory controllers, and DRAM address buses, effectively eliminating performance interference between different applications. 
Each GPU instance contains at least one GPU Processing Cluster (GPC) along with a designated portion of the GPU's memory. 
The current setup of MIG can support up to seven distinct instances, offering predefined configurations including 1g, 2g, 3g, 4g, and 7g, where `g' indicates a portion of the total GPU compute resources. For instance, the smallest configuration available is 1g.5gb, providing 1/7 of the Streaming Multiprocessors (SMs) and 5\,GB of GPU memory. However, configurations for 5g and 6g are not available.

\begin{figure}[tb]
	\centering
\includegraphics[width=0.46\textwidth]{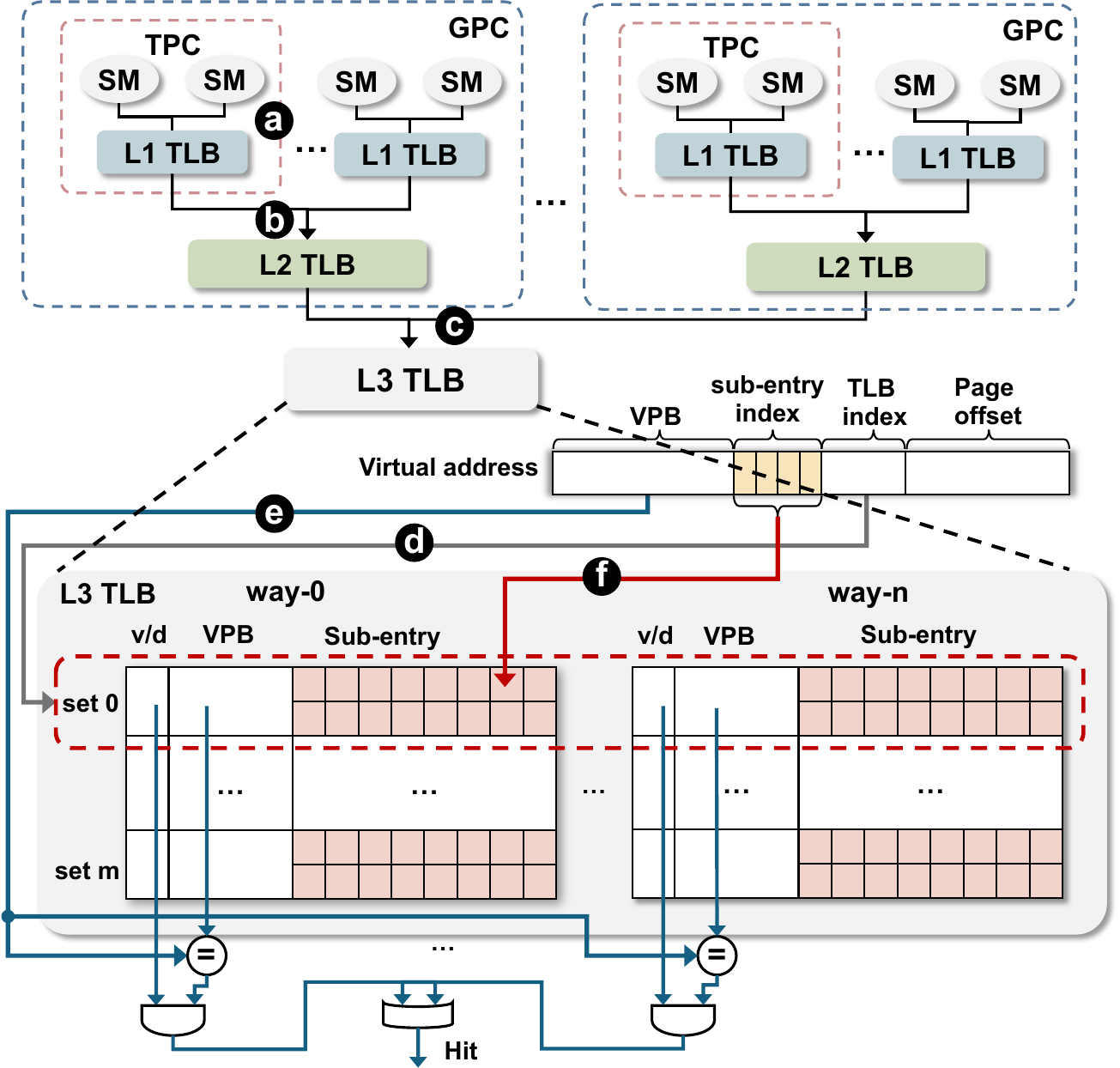}
\vspace{-8pt}
\caption{TLB structure and address translation process in A100. }

\label{fig:baseline_arch}
\vspace{-16pt}
\end{figure}

The TLB organization in MIG is shown in Figure~\ref{fig:baseline_arch}. Specifically, MIG partitions the L1 and L2 TLBs along with the GPCs: the L1 TLB is shared between the two SMs within each Texture Processing Cluster (TPC), and the L2 TLB is shared across the SMs of a GPC. However, interestingly, prior work \cite{zhang2023t} reveals that the L3 TLB in today's advanced GPUs (e.g., NVIDIA's Ampere generations) remains shared across all instances in MIG-supported GPUs.
This sharing indicates that despite MIG's comprehensive approach to partitioning, the last-level TLB still lacks the isolation necessary to prevent contention across different GPU instances.

\subsection{Address Translation in MIG}
\label{subsec:addresstranslation_in_mig}
Figure~\ref{fig:baseline_arch} also illustrates the address translation process in MIG. 
Upon a memory request, the L1 cache is first checked. The L1 TLB lookups are performed upon an L1 cache miss (\circled{a}). If the request misses in the L1 TLB, it first checks the L1 Miss Status Holding Register (MSHR) to coalesce the same requests and the outstanding request is sent to the L2 TLB for lookup (\circled{b}). Similarly, requests missing in the L2 TLB are sent to the L3 TLB (\circled{c}), and requests that miss the L3 TLB are further sent to the GPU memory management unit (GMMU) to perform page table walks. If the page table walk fails, a local page fault is generated and propagated to the host CPU to resolve. It then initiates the target data transfer and updates TLBs, caches, and page tables. Finally, the address translation request is replayed after resolving the page fault.

\vspace{1mm}
\noindent{\bf TLB sub-entries:}
Traditionally, each TLB entry would directly map one virtual page to one physical page. This is a straightforward, one-to-one relationship: each entry in the TLB represents a single page of memory, as typically done in L1 TLBs in the latest NVIDIA GPUs (e.g., NVIDIA's Ampere generations). However, these GPUs organize their L2 and L3 TLB entries differently to increase TLB reach~\cite{zhang2023t}.
Specifically, each of these entries contains 16 sub-entries, which directly map to the address translations for 16 pages. These pages can be either 64\,KB or 2\,MB in size, and all of them fall within an aligned range of either 1\,MB or 32\,MB in size, respectively. 
That means each sub-entry in a TLB entry has a one-to-one relationship with a single page.
Note that, in the sub-entry setting, if any TLB entry is evicted, all the 16 sub-entries associated with that TLB entry are zeroed.

Address translation in a sub-entry TLB proceeds as follows.
The virtual address of memory access is partitioned into a virtual page number (VPN) and a page offset. 
The lower bits of the VPN are further divided into a TLB index and a sub-entry index, and the higher bits of the VPN serve as the virtual page base (VPB). 
During a TLB lookup, the TLB index is first used to identify the corresponding set (\circled{d}). Then, the VPB from the virtual address is compared with the VPB within the set to check for a hit or miss (\circled{e}). If there is an entry hit, the process further checks the sub-entry index to determine if the specific sub-entry is present in the TLB entry (\circled{f}). A non-zero sub-entry indicates a TLB hit.
Conversely, a zero sub-entry or no matching VPB results in a TLB miss, triggering a page table walk.
When a valid translation for a virtual address is found and
if this virtual address is within the range covered by an existing TLB entry, the translation is added to the appropriate sub-entry slot.  If no existing TLB entry covers this address range, a new entry is created. This involves evicting the least recently used (LRU) entry along with zeroing all of its 16 sub-entries. The new translation is then inserted into the corresponding sub-entry slot of the newly established TLB entry.

\section{Methodology}
\label{sec:motivation}

\subsection{Baseline Configuration}
\label{ssec:3.1}

\vspace{-4pt}
\begin{table}[ht]
  \centering
  \scriptsize
    \vspace{-8pt}
  \caption{Baseline multi-instance GPU configuration.}
  \vspace{-4pt}
  \begin{tabular}{|l|l|}
    \hline
    \textbf{Module} & \textbf{Configuration}\\
    \hline
    \hline
    SM & 1 GHz, 108 in total \\
    \hline
    DRAM & 5\,GB per slice\\
    \hline
    L1 D-cache &  64\,KB, 2-way set associative \\
    \hline
    L1 I-cache &  32\,KB, 2-way set associative \\
    \hline
    L2 cache & 2\,MB per slice, 8-way set associative \\
    \hline
    L1 TLB & 16 entries, 16-way, 1-cycle lookup latency, \\
    & TPC shared, LRU replacement policy\\
    \hline
    L2 TLB & 128 entries, 8-way, 16 sub-entries per entry, \\  
           & 10-cycle lookup latency, GPC shared, \\
           & LRU replacement policy \\
    \hline
    L3 TLB & 1024 entries, 8-way, 16 sub-entries per entry, \\
           & 40-cycle lookup latency, GPU shared, \\
           & LRU replacement policy\\
    \hline
    Page table walk & 8 page table walkers, GPC shared, \\
    & 100-cycle latency per level~\cite{8416827, 7482091, pratheekimproving} \\
    \hline
    Page walk cache & 128 entries shared across page table walkers~\cite{pratheekimproving} \\
    \hline

  \end{tabular}
  \label{table:config}

\end{table}
\vspace{-4pt}

We use MGPUSim~\cite{sun19mgpusim} throughout the paper. To model multi-instance GPU, we substantially modified MGPUSim by adding (i) different cache, memory, and SM configurations for different instance sizes, (ii) a shared L3 TLB (with sub-entries) and sub-entries for the L2 TLB, and (iii) GMMUs for each instance, including page walk cache, page walk queue, page table walk thread, and the page table. Note that the exact latency of a page walk depends on whether it hits the page walk cache and whether it needs to wait for an available page walk thread in the page walk queue. These processes are all faithfully modeled in the simulator.
In this paper, we focus on a GPU partitioned into instances of sizes 3g, 2g, and 2g. The number of SMs, cache size, and memory size are partitioned proportionally based on each instance size. Each instance runs a single application. 
Our approach also applies to various combinations of instance sizes and we provide a sensitivity study with altering instance sizes in Section~\ref{ssec:sensitivity}. The baseline configuration is listed in Table~\ref{table:config}.  The page size is set to 64\,KB as the MIG default configuration.

\subsection{Applications}

We use 8 applications from the Polybench~\cite{polybench}, SHOC~\cite{10.1145/1735688.1735702}, Hetero-Mark~\cite{7581262}, AMDAPPSDK~\cite{amdapp}, Rodinia~\cite{5306797}, and DNN Mark~\cite{dong2017dnnmark} benchmark suites, which are representative real-world applications. The details of the applications are listed in Table~\ref{table:app}. 
These applications have different computation intensities. For example, {\tt FFT} and {\tt CONV} are compute-intensive and heavily use floating-point operations, whereas {\tt BICG}  and {\tt ATAX} involve memory-intensive operations. 
The applications also cover a wide range of access patterns. Specifically, in the stream access pattern, data is accessed sequentially, offering good locality and predictability. In contrast, the stride access pattern, shown in operations like matrix transpose, involves accessing data at a constant stride, leading to non-sequential memory accesses. For example, in {\tt MT}, accessing elements column-wise in a row-major stored matrix or vice versa involves memory accesses with a stride equal to the number of rows or columns.
In the dependent access pattern, certain data is accessed depending on the computation results of previous elements, such as in {\tt NW}, where each cell's computation in the scoring matrix depends on the values of its neighboring cells. In the block access pattern, data is accessed in blocks or chunks. For example, in {\tt ST}, data is divided into blocks that fit into the cache, allowing for efficient computation of the convolution operation over each block.

\begin{table}[t!]
  \centering
  \scriptsize
  \setlength\tabcolsep{1.8pt} 
 \caption{List of applications. }
 \vspace{-4pt}
  \begin{tabular}{|l|l|l|r|r|l|}
    \hline
    \textbf{Abbr.\hspace{-1pt}} & \textbf{Application} &  \textbf{\makecell{Benchmark\\Suite}} & \textbf{\makecell{Instruction\\ Count}} & \textbf{\makecell{\hspace{-1.5pt}L2 TLB\hspace{-1.5pt} \\ MPKI}} &\textbf{\makecell{Access\\Pattern} }
    \\
    \hline
    \hline
    {\tt ATAX} & \makecell[l]{Matrix Transpose and\\  Vector Multiplication} & Polybench & 328,441,844 & 204.7  & Stream, Stride\\
    \hline
    {\tt BICG} & \makecell[l]{Sub Kernel of BiCG-\\ Stab Linear Solver} & Polybench & 321,758,896 & 208.9  & Stream, Stride\\
    \hline
     {\tt FFT} & \hspace{-1.5pt}Fast Fourier Transform\hspace{-1.5pt} & SHOC &409,534,464 & 0.5 & \makecell[l]{Stream, Stride} \\
     \hline
     {\tt ST} & Stencil 2D & SHOC & 59,289,600 & 21.9 &  Stream, Block\\
     \hline
    {\tt FIR} & Finite Impulse Resp. & Hetero-Mark & 192,675,840 & 0.3  & Stream\\
    \hline
    {\tt MT} & Matrix Transpose & \hspace{-1.5pt}AMDAPPSDK\hspace{-1.5pt} & 9,564,256 & 205.0 &  Stride\\
    \hline
    {\tt NW} & Needleman–Wunsch & Rodinia & 87,909,120 & 38.4 & \makecell[l]{Stream,\\  Dependent}  \\
    \hline
    {\tt CONV} & Convolution 2D & DNN-Mark & \hspace{-1.5pt}2,629,570,744\hspace{-1.5pt}  & 1.9 &  Stream, Stride\\
   
    \hline
  \end{tabular}
   \vspace{-8pt}
  \label{table:app}

\end{table}

\begin{scriptsize}
\begin{table}[t]
\scriptsize
  \centering
  \caption{Multi-tenancy workloads.}
  \vspace{-8pt}
  \begin{tabular}{|l|l|l|l|}
    \hline
    \textbf{Abbr.} & \textbf{Workload} & \textbf{Applications} & \textbf{Category}\\
    \hline
    \hline
    W1 & workload1 & {\tt MT}, {\tt ATAX}, {\tt BICG} & HHH\\
    \hline
    W2 & workload2 & {\tt MT}, {\tt ATAX}, {\tt ST} & HHM\\
    \hline
    W3 & workload3 & {\tt MT}, {\tt NW}, {\tt ST} & HMM\\
    \hline
    W4 & workload4 & {\tt MT\_s}, {\tt ST\_s}, {\tt FIR} & HML\\
    \hline
    W5 & workload5 & {\tt MT\_s}, {\tt FFT}, {\tt FIR} & HLL\\
    \hline
    W6 & workload6 & {\tt NW}, {\tt CONV}, {\tt ST\_s} & MMM \\
    \hline
    W7 & workload7 & {\tt ST\_s}, {\tt NW}, {\tt FFT} & MML \\
    \hline
    W8 & workload8 &  {\tt ST\_s}, {\tt FIR}, {\tt FFT} & MLL \\
    \hline
    W9 & workload9 & {\tt FFT}, {\tt FFT}, {\tt FIR} & LLL\\
    \hline
    
  \end{tabular}
  \label{table:mul-appWorkload}
\vspace{-8pt}
\end{table}
\end{scriptsize}

To study multi-instance execution, we use the applications listed in Table~\ref{table:app} to form multi-application workloads. 
We also include applications with their smaller input size (indicated as {\tt ApplicationName\_s}) to balance the application execution times within the workload.
Table~\ref{table:mul-appWorkload} shows the nine workloads, each consisting of three applications. 
The workloads are formed by analyzing the L3 TLB access intensity of each application. Specifically, we measure each application's misses-per-kilo-instructions (MPKI) of the address translations at L2 TLB. 
Applications are then grouped into three categories based on their L2 TLB MPKI values: Low ($L$, MPKI$<$1), Medium ($M$, 1$<$MPKI$<$100), and High ($H$, MPKI$>$100). Accordingly, workloads are formed representing various combinations of these categories, including $HHH$, $HHM$, $HMM$, $HML$, $HLL$, $MMM$, $MML$, $MLL$ and $LLL$.
Given the possibility of some applications finishing earlier than others during simultaneous execution, we adopt the same strategies as previous studies to ensure continuous TLB contention~\cite{10.1145/1555754.1555778, pratheekimproving, li2021improving}. 
That is, applications that are completed early are re-run until the completion of the longest-running application within the workload. The statistical data is gathered only during the initial complete run of each application within any given workload.

\section{Quantitative Analysis of MIG Multi-Tenancy}
\label{sec:motivation}

\subsection{Overall Performance Characteristics}

\label{ssec:overallPerfStudy}

\begin{figure}[!h]
	\centering
\includegraphics[width=0.46\textwidth]{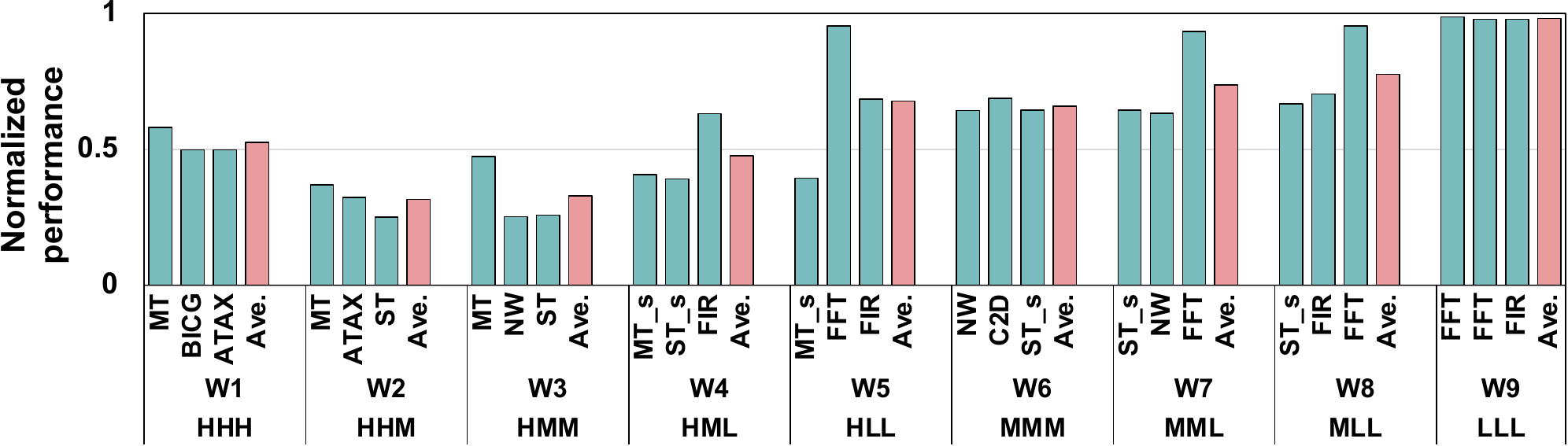}
\vspace{-8pt}
\caption{The normalized performance of each application within the workload. }
\label{fig:overall_perf}
\vspace{-2pt}
\end{figure}


In a MIG-enabled GPU, it has a key source of contention under multi-tenant execution, namely the shared L3 TLB. To quantify the performance impact of interference and contention at the L3 TLB, we study normalized performance of individual applications within workloads and the average performance of the nine workloads in Table~\ref{table:mul-appWorkload} as shown in Figure~\ref{fig:overall_perf}. Specifically, the normalized performance here is the performance of an application executed in conjunction with other applications, normalized to the performance of running alone. The average performance is calculated as the harmonic mean of normalized performance for all applications within a workload. Note that, when an application runs alone, it uses the same instance size but gets exclusive use of the full L3 TLB capacity.
One can make the following observations. First, L3 TLB contention compromises the performance of individual applications. In W9, each application experiences a negligible performance decrement. Conversely, in W1, there is an average performance drop of 48\%. Second, the performance degradation varies among different applications within the same workload. This variance is particularly significant in applications with higher MPKI values. As shown in W7, where the performance of the {\tt FFT}, with a low MPKI of 0.5, drops by 6.6\%, in contrast to the {\tt NW}, which suffers a substantial 36.7\% decrease with a medium MPKI of 38.4.
This is because applications with higher MPKI values are more sensitive to TLB misses due to limited latency hiding through context switching or other parallel threads. 
Third, the performance degradation of the same application can vary depending on the specific co-runners. Taking {\tt ST\_s} as an example, its performance drops by 61\% in W4 but only by 34\% in W8. This is due to the co-running applications having a higher MPKI in W4 than those in W8, which leads to more severe L3 TLB contention.

\begin{figure}[!t]
	\centering
\includegraphics[width=0.46\textwidth]{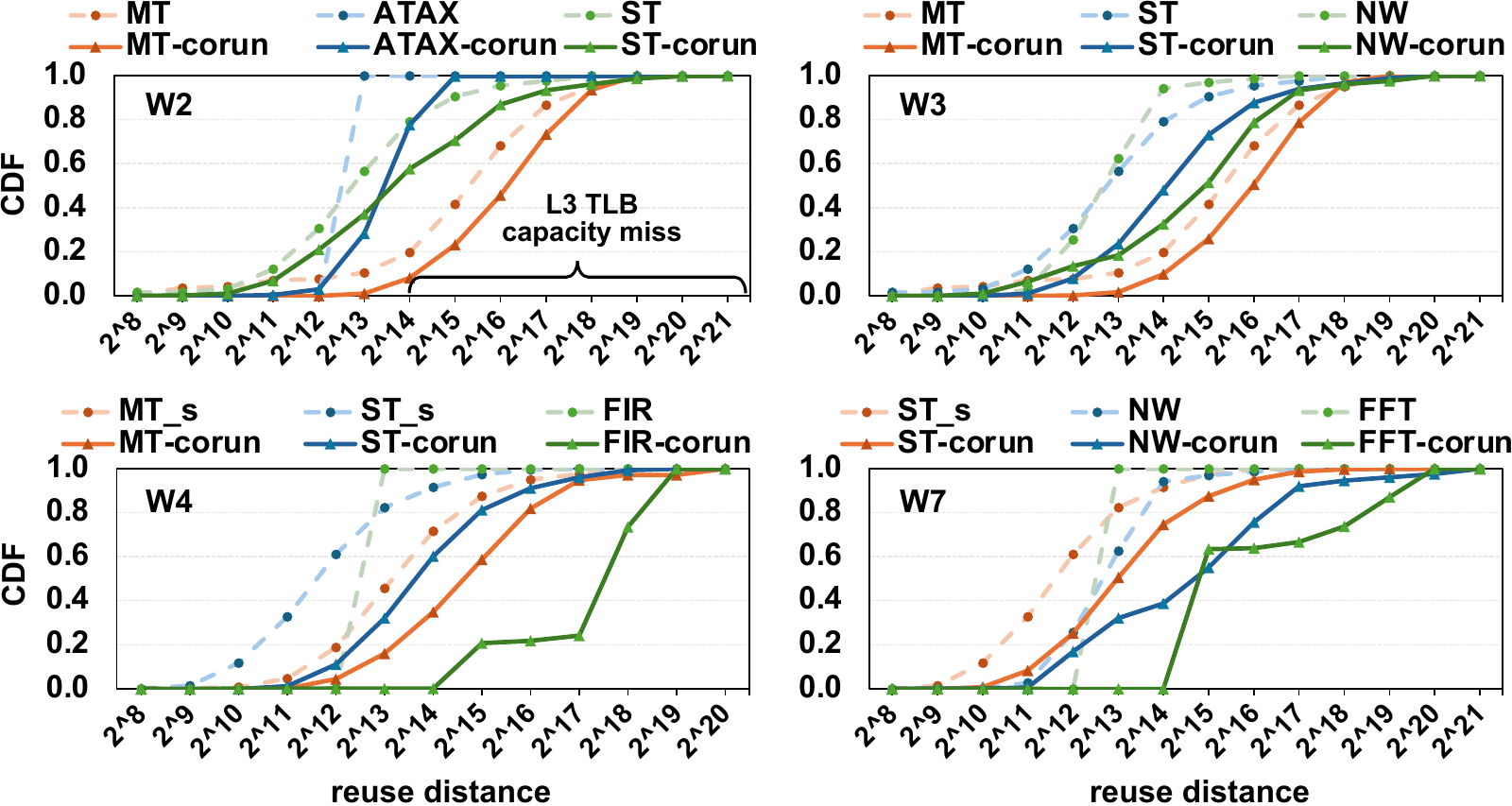}
\vspace{-8pt}
\caption{CDF of translation reuse distances at the L3 TLB. }
\vspace{-12pt}
\label{fig:reuse}
\end{figure}

We further investigate the reuse distance of translations for multi-tenancy. The reuse distance is defined as the unique translation count between two accesses to the same translation from the same instance. In multi-tenant execution, we calculate reuse distance considering the application process ID to differentiate the reuses from different applications within the workload mix.
Figure~\ref{fig:reuse} presents the Cumulative Distribution Function (CDF) of the translation reuse distances of four workloads with representative MPKI mix, i.e., HHM, HMM, HML, and MML. For comparison, we also show the reuse distance for each application running alone, depicted in light dotted lines. 
We observe that some applications (e.g., {\tt NW}, {\tt FFT} and {\tt FIR}) show very different reuse distances when they execute concurrently with others versus running alone. For example, in its single-run of {\tt NW}, 94.2\% translation reuses are less than the L3 TLB capacity (i.e., 16,384 sub-entries), indicating a higher possibility of these reuses being accommodated within the TLB. However, in W3, only 32.7\% of the reuses in {\tt NW} are within the TLB capacity.
This is because in W3, {\tt ST} and {\tt MT} have high/medium MPKIs, and they generate a large number of translation requests to the L3 TLB. Therefore, the reuse distance of {\tt NW} is extended. 
For applications such as {\tt ST\_s} in W7, its reuse distance shows relatively little change compared to its isolated run. This is because it generates intensive translation requests that miss in the L2 TLB and therefore consume a considerable portion of L3 TLB entries; at the same time, its co-located application {\tt FFT} has a lower MPKI, thereby generating less contention for TLB resources.   
We also marked the L3 TLB capacity in the figure. It is observed that for applications with severe contention  (e.g., {\tt MT}), more than 80\% of the translation reuses miss in the L3 TLB.

\subsection{Sub-Entry Utilization Characterization}
\label{ssec: sub-utilization}

\begin{figure}[!t]
	\centering
\includegraphics[width=0.42\textwidth]{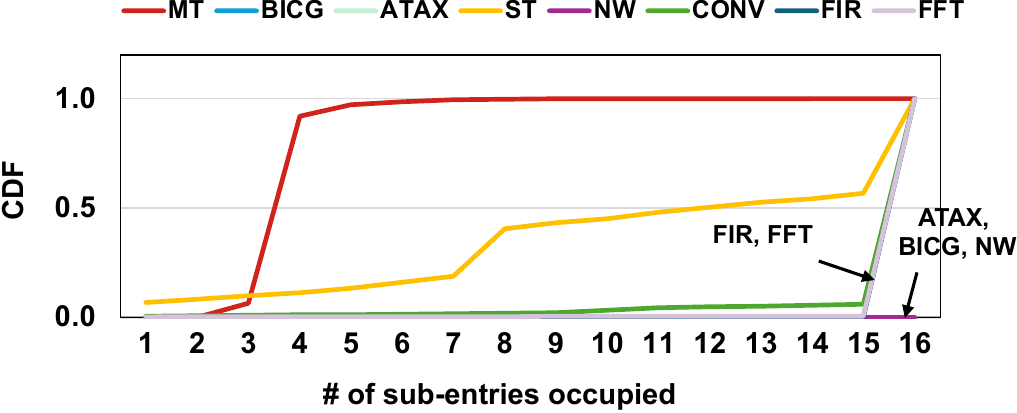}
\vspace{-10pt}
\caption{CDF of TLB sub-entry utilization when running in isolation. }
\label{fig:util_single}
\end{figure}

Recall that, 
when a TLB entry is evicted, the sub-entries within the TLB entry are also evicted. This design is beneficial for scenarios where memory accesses exhibit a contiguous or linear pattern. In such cases, most sub-entries can be utilized effectively before any TLB entry is evicted, thereby maximizing the efficiency of the TLB. However, in situations where memory access patterns become non-contiguous, particularly in workloads with sparse or irregular memory access patterns, some sub-entries might remain unused at the time of eviction. We therefore study the utilization of sub-entries of each application when it is evicted.  
Figure~\ref{fig:util_single} shows the CDF of sub-entry utilization of all applications listed in Table~\ref{table:app} running in isolation.
One can observe that applications with stream access patterns, such as {\tt FIR} and {\tt FFT}, tend to make full use of TLB sub-entries before eviction due to their sequential access nature. In contrast, the application {\tt MT} exhibits low sub-entry utilization (most TLB entries evicted with only four sub-entry occupied). This is because {\tt MT} has stride access patterns, where accesses do not align well with the contiguous page mappings of the sub-entries. Moreover, application {\tt ST}, which exhibits a block access pattern along with a stream pattern, shows nearly 50\% of the TLB entries are evicted when only half of the sub-entries are utilized. This is because of the mixed nature of its memory accesses, i.e., sequential within blocks but non-contiguous between them. Note that the memory footprints of applications {\tt ATAX}, {\tt BICG}, and {\tt NW} can fit in the address coverage range of L3 TLB, therefore no eviction is observed when they are running alone.

\begin{figure}[!t]
	\centering
\includegraphics[width=0.5\textwidth]{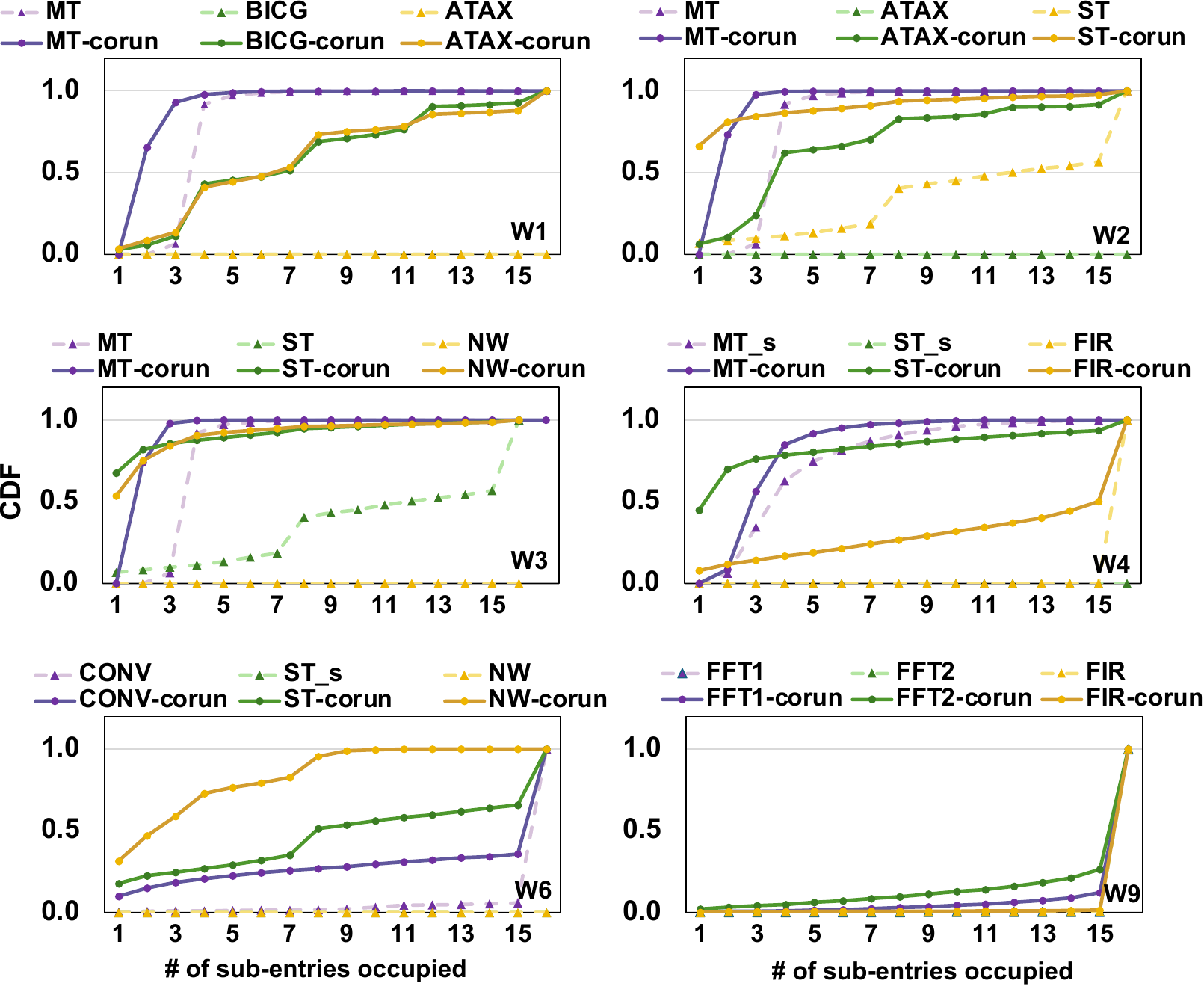}
\vspace{-18pt}
\caption{CDF of TLB sub-entry utilization under co-running. }
\vspace{-12pt}
\label{fig:subentry_util}
\end{figure}

We further analyze the contention and interference impact on sub-entry utilization when co-running applications. Figure~\ref{fig:subentry_util} presents the sub-entry utilization of six workloads with representative MPKI mix, i.e.,
HHH, HHM, HMM, HML, MMM, and LLL. The darker solid lines in the figure represent the sub-entry utilization of each application when co-running; we also show the sub-entry utilization when applications run individually in lighter dot lines. One can make the following observations. First, all applications within workload categories, except LLL, show substantially less utilization of sub-entries when a TLB entry is evicted compared to applications that are run in isolation. For example, in the application {\tt ATAX} in W1, 73.4\% of its TLB entries are evicted when less than half of the sub-entries are used, while no evictions during its individual run. Similarly, for application {\tt ST} in W2, 66.3\% of its TLB entries are evicted with only one sub-entry used, whereas 43.4\% of its TLB entries are evicted with fully occupied sub-entries when run in isolation. More severe underutilization is observed for workloads with a larger MPKI mix. Second, the same application has very different utilization in different workloads. For example, {\tt ST\_s} in W4 has 69.8\% of its TLB entries with just two sub-entries used at the time of eviction. In contrast, in W6, {\tt ST\_s} shows only 22.5\% of its entries evicted with two sub-entries used. This is because in W4, the co-running application {\tt MT\_s} has high MPKI, which leads to a greater number of translation requests to the L3 TLB, causing {\tt ST\_s} to suffer from more frequent evictions before the sub-entry is fully utilized due to increased contention.

\vspace{1mm}
\noindent\textbf{Does prompting sub-entry to regular TLB entry solve the problem?} A straightforward approach to enhancing sub-entry utilization would be to convert sub-entries into regular TLB entries, thus eliminating the 1\,MB virtual address range alignment for each TLB entry and allowing any address to utilize these sub-entries. However, such an expansion would result in a significant hardware cost. In the baseline, each way uses one comparator to match the incoming address with stored tags. Requests that fall within a specific TLB entry range directly map to the corresponding sub-entry, simplifying comparisons. However, allowing any address to use a sub-entry would require each of the 16 sub-entries in a TLB entry to have its comparator.  Since each TLB entry is associated with 16 sub-entries, this would increase the number of comparators by 16 for each way. We use CACTI~\cite{cacti} to estimate the TLB size: under this design, the TLB size is 17.2$\times$ of the baseline. This increase is impractical considering the constraints on GPU die size. Therefore, it is important to explore a more efficient and cost-effective approach to optimize TLB sub-entry utilization without excessively increasing its size.


\section{Sub-Entry Sharing-Aware TLB}
Our goal in this paper is to improve the MIG-enabled GPU TLB hit rates, thereby boosting the performance of multi-tenancy execution. While contention for a shared resource is inevitable in environments where resources are limited, our analysis in the previous section has revealed opportunities to mitigate such contention's ill effects by optimizing the utilization of TLB sub-entries.

To this end, we propose \ourdesign, a hardware-supported TLB sub-entry sharing mechanism that allows multiple base addresses to share a TLB entry of 16 sub-entries dynamically. Our approach organizes sub-entries into multiple groups, allocating each group to one base address for usage. However, implementing an effective and efficient dynamic sub-entry sharing mechanism is non-trivial and faces several challenges. First, reducing the number of sub-entries allocated to each base address changes the direct mapping from the original design.  It is important to resolve any resulting conflicts while maintaining the correctness of address translation lookup.
Second, it is important to select appropriate base addresses for sharing and determine when to share such that the utilization can be maximized and minimize the performance impact compared to the original sub-entry capacity. Third, enabling sub-entry sharing, the TLB lookup and insertion procedure should not significantly be increased compared to the baseline. Finally, the proposed TLB sub-entry sharing should involve minimum hardware overheads, offering a cost-effective and scalable alternative to merely enlarging the TLB size. 


\subsection{Sub-Entry Sharing-Aware TLB Format}

\begin{figure}[!t]
	\centering
\includegraphics[width=0.47\textwidth]{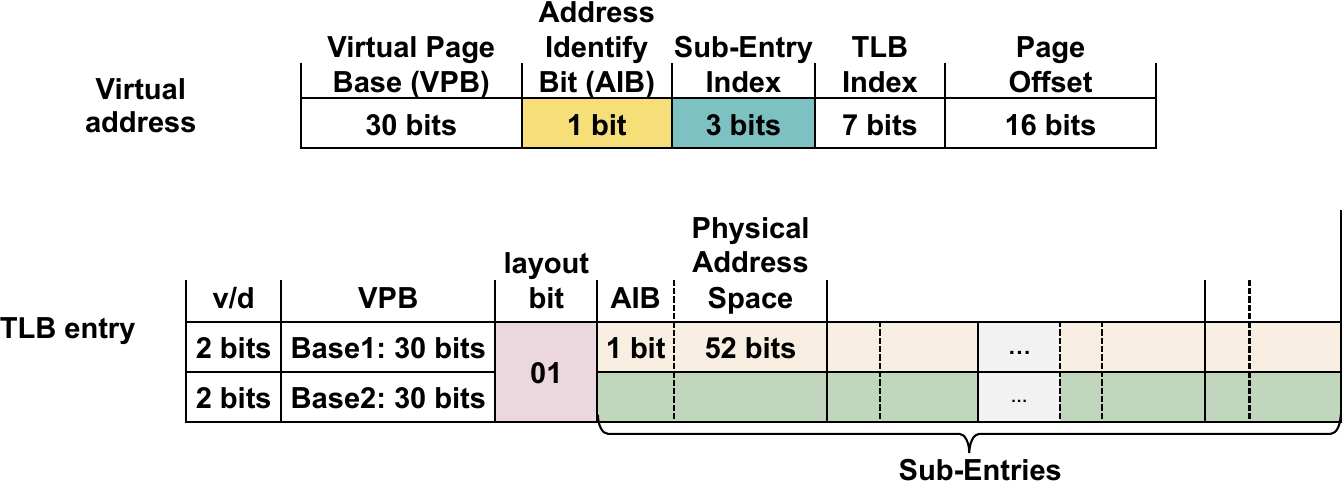}
\vspace{-4pt}
\caption{Format of the virtual address and contents of a sharing-aware TLB entry in sequential layout. }
\label{fig:subentry_format_0}
\vspace{-8pt}
\end{figure}

\begin{figure}[!t]
	\centering
\includegraphics[width=0.47\textwidth]{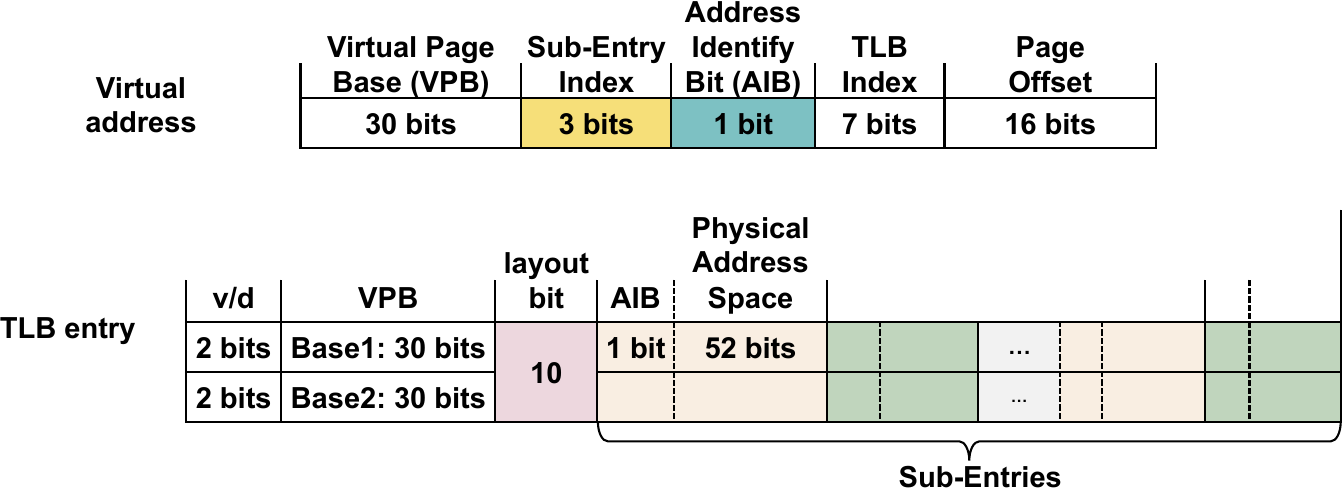}
\vspace{-4pt}
\caption{Format of the virtual address and contents of a sharing-aware TLB entry in stride layout. }
\vspace{-12pt}
\label{fig:subentry_format_1}
\end{figure}

Figure~\ref{fig:subentry_format_0} depicts the format of virtual addresses and the content of a sharing-aware TLB entry. Specifically, the original 4-bit sub-entry index is split into an $n$-bit sub-entry index and a ($4-n$)-bit Address Identify Bit (AIB). The value of $n$ depends on how many base addresses are sharing one TLB entry. In the current design, we allow each original TLB entry to support two base addresses, and each address can occupy eight sub-entries (i.e., $n=3$). This pre-determined value is based on our characterization analysis presented in Section~\ref{ssec: sub-utilization}, where we found over half of the TLB entries were evicted while less than half of their 16 sub-entries were utilized. The shared TLB entry also needs additional bits to maintain the metadata (e.g., valid/dirty bits) for each base address separately. 
Since each base address is limited to using 8 sub-entries in our design, the absence of a direct one-to-one mapping within a 1\,MB alignment could lead to conflicts for sub-entries with identical index bits. To address this, our approach dictates that if a sub-entry is already in use and a new request arrives with the same sub-entry index bit, the new request will replace the existing one. Consequently, at any given moment when a TLB entry is shared, only one address translation with a particular sub-entry index bit can be presented in a TLB entry. The Address Identifier Bit (AIB) becomes essential here, serving to identify which address is currently using the sub-entry.

Because of the diverse access patterns exhibited by applications, e.g., stream versus stride patterns, we introduce a flexible method that dynamically allocates sub-entries to base addresses based on the usage patterns of sub-entries.
Specifically, for scenarios where sub-entries are occupied sequentially, we allocate the first half of the sub-entries to the first base address and the second half to the second base address. In this case, the last three bits of the sub-entry index are used to identify positions within each base's allocated sub-entries. The first bit of the sub-entry index acts as the Address Identifier Bit (AIB) (shown in Figure~\ref{fig:subentry_format_0}). Alternatively, if the occupied sub-entries show stride access patterns, the sub-entries are interleavedly allocated between the two base addresses according to the stride size. In our approach, we pre-defined stride size as 1. That is, the first base address is assigned to sub-entries with even indices, whereas the second base address is allocated to those with odd indices. In this case, the first three bits of the sub-entry index are used to determine the location of sub-entries (shown in Figure~\ref{fig:subentry_format_1}). 
These sub-entry layout strategies are recorded in layout-bit (initially set to `00', indicating non-shared status). 
When inserting the new base address, the choice between sequential or stride layout depends on the \textit{current occupancy pattern} of sub-entries: a consecutive pattern triggers the sequential layout (layout bit set to `01'), whereas a non-consecutive pattern activates stride layout (layout bit set to `10'). The layout bit then determines which index bits are used during a lookup.

Note that choosing between different numbers of shared base addresses leads to a trade-off between sub-entry utilization and hardware overhead. More shared base addresses increase sub-entry utilization but require more bits to be stored in the TLBs and more cycles to compare each base address. On the other hand, fewer shared base addresses reduce hardware overhead and lookup latency but lead to lower sub-entry utilization. We provide sensitivity results with different numbers of shared base addresses in Section~\ref{ssec:sensitivity}.

\subsection{TLB Lookup and Insertion Process}

\noindent\textbf{When to share?} In our sub-entry sharing-aware TLB architecture, sub-entry sharing is allowed under specific conditions to optimize utilization. Initially, the TLB works as the default baseline, with each TLB entry independently managed until all entries (ways) within the TLB set are allocated. At that point when a new address arrives, instead of proceeding with a Least Recently Used (LRU) entry eviction, we first check how many sub-entries are actually being used in all entries of that set. An entry is considered eligible for sharing if it meets the following criteria: (i) less than eight sub-entries are utilized, and (ii) only one base address is currently occupying the entry. If multiple entries fit these criteria, we prefer to pair the incoming base address with an existing entry from the same process because access patterns within the same process tend to be similar. 
If no matching process is found, we choose the candidate where the current sub-entry utilization is the lowest. Only if no entries meet these conditions for sharing do we fall back to the baseline approach of evicting the least recently used entry and inserting the new one. 

\ourdesign~also supports dynamic shifts between the shared and non-share status. The shared TLB entry can still revert to the non-shared status. Specifically, when a TLB entry, currently shared between two base addresses, reaches a state where all 8 sub-entries allocated to one base are fully utilized, it indicates the potential for an increase in demand for this process. Upon the arrival of a new request that cannot be accommodated due to fully utilized sub-entries for its corresponding base address, the shared TLB entry will be reverted to being exclusively used by one base with increasing demand. The other base and its associated sub-entries are evicted from the TLB entry.  The TLB entry status is updated, which involves resetting the layout bit, the metadata for the second base, and reorganizing the sub-entries based on the 4-bit sub-entry index.

\begin{figure}[!t]
	\centering
\includegraphics[width=0.48\textwidth]{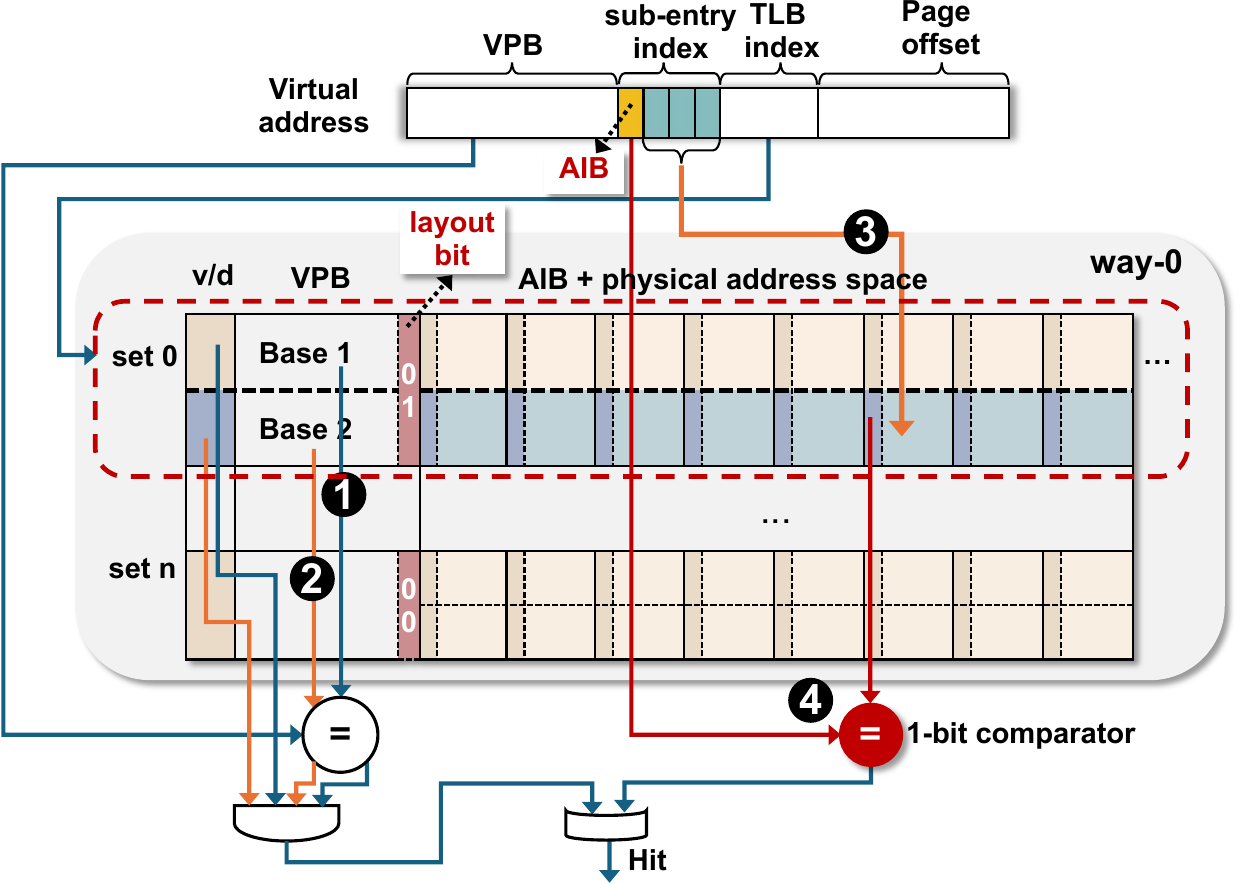}
\vspace{-6pt}
\caption{TLB lookup process in \ourdesign. }
\vspace{-12pt}
\label{fig:our_design}
\end{figure}

\vspace{1mm}
\noindent\textbf{TLB Lookup:} 
Figure~\ref{fig:our_design} shows the TLB lookup procedure, which is provided in Algorithm~\ref{algo:look}. 
Specifically, when a translation request arrives at the L3 TLB, it first identifies the corresponding set. Each entry in that set is compared in parallel with the request's VPB. Two scenarios can happen. First, the entry is non-shared with only one base address, indicated by the layout bit set to `00'. The 4-bit sub-entry index is then used to locate the corresponding sub-entry. Second, if an entry contains two base addresses, these addresses are checked sequentially (\circled{1}, \circled{2}). Upon identifying a matching entry, the layout bit is checked to determine which bits should be used to index the corresponding sub-entry (\circled{3}). If the layout bit is set to `01', the last three bits of the sub-entry index are employed to locate the specific sub-entry. On the other hand, if the layout bit is set to `10', the first three bits will be used. Once locating the sub-entry, the Address Identify Bit (AIB) stored in the sub-entry is compared with the request's AIB (\circled{4}). A matching AIB indicates a TLB hit, and the PFN stored in the sub-entry is concatenated with the page offset to form the requested physical address. An AIB miss (also TLB miss) is handled the same way as in the baseline which involves sending the request to GMMU for a page table walk.
Note that the sequential lookup latency is twice that of the baseline, and these latency overheads are included in our evaluation.

\begin{algorithm}[!t]
\scriptsize
\DontPrintSemicolon
\caption{TLB Lookup with Sub-Entry Sharing.}
\LinesNumbered 
\label{algo:look}

\textbf{/* Lookup () */}

Request arrives L3\_TLB\;
Compare each entry in set with request's VPB in parallel\;
\eIf{non-shared TLB entry}{
    Use 4-bit of sub-entry index\;
     \eIf{find matched sub-entry}{
     TLB hit, respond with PFN concatenated with page offset \;
     }{
     TLB miss, send request to GMMU for page table walk \;
     }
}{
    Check base addresses sequentially \;
    \eIf{find matched base address}{
        Check layout bit \;
        \eIf{layout bit equals `01'}{
                Use last three bits of sub-entry index\;
        }{
            Use first three bits of sub-entry index\;
        }
        Locate sub-entry and compare AIB with request's AIB \;
        \eIf{AIB matches}{
            TLB hit \;
        }{
            TLB miss \;
        }
    }{
        TLB miss \;
    }
}
\end{algorithm}

\begin{algorithm}[!h]
\scriptsize
\DontPrintSemicolon
\caption{TLB Insertion with Sub-Entry Sharing.}
\LinesNumbered 
\label{algo:Insert}

\textbf{/* Insert ()*/}

Find the TLB set for the virtual address.\;
\tcc{Scenario 1: Base address hit}
\eIf{address matches an existing base}{
    \uIf{entry is shared}{
        Use the layout bit to find the sub-entry.\;
        \If{base's sub-entries are full}{
            Make entry non-shared.\;
            Reset the layout bit. \; 
            Reorganize sub-entries.\;
        }
        \Else{
            Insert into sub-entry. Evict the original if needed.\;
        }
    }
    \Else{
        Insert translation with 4-bit index.\;
    }
}{
    \tcc{Scenario 2: Miss all base addresses}
    \If{there is an available entry in the set}{
        Insert new base address into first vacant entry\;
    }
    \Else{
        \If{conditions for sub-entry sharing are met}{
        Check access pattern of sub-entries\;
        \uIf{sub-entries are continuously occupied}{
            Apply sequential layout; set layout bit to `01'\;
        }
        \Else{
            Apply stride layout; set layout bit to `10'\;
        }
        \uIf{sub-entry is already occupied by the original base}{
            Try to relocate the original entry\;
            \If{alternative sub-entry is also occupied}{
                Evict the original entry to accommodate new translation\;
            }
        }
        \Else{
            Insert new address translation with determined layout\;
        }
    }
    \Else{
        Evict least recently used (LRU) entry and insert new address\;
    }
}
}
\end{algorithm}

\vspace{1mm}
\noindent\textbf{TLB Insertion:}
The insertion algorithm is shown in Algorithm~\ref{algo:Insert}.
When a new address needs to be inserted, the index bits of the virtual page number are used to determine the
set in the TLB. Two scenarios can happen depending on whether the address matches an existing base address in the identified set. 
In the first scenario, if the base address hits, the following process depends on the shared status of the TLB entry. For a non-shared TLB entry (layout bit `00'), the address translation is inserted into its corresponding sub-entry using the complete 4-bit sub-entry index. 
On the other hand, if the TLB entry is shared by two base addresses,  the layout bit determines whether the last or first three bits of the index are used to locate the sub-entry. 
In a situation where all sub-entries for the inserted base address are full, it triggers a shift from a shared to a non-shared status. That is, the metadata and sub-entries associated with the other shared base address are evicted and the layout bit is reset to `00'. The sub-entries will be relocated using the 4-bit sub-entry index. Instead, if the sub-entries of the inserted base address are not full, the incoming translation is inserted into the sub-entry as indicated by the 3-bit sub-entry index; if the target location is already occupied, the original translation is removed.

In the second scenario, it misses all base addresses. 
If there is an available entry within the set, the new base address is inserted into this first vacant entry. Otherwise,  the conditions for sub-entry sharing are evaluated, as previously discussed.
If an entry is selected for sharing with the new base, the access pattern of the current entry is checked to determine how to organize the shared sub-entries (i.e., sub-entry layout).  
Specifically, if the sub-entries are occupied continuously without any gaps, it is classified as a sequential pattern. For such cases, the sequential layout will be applied to the sub-entries, and the layout bit is set to `10' (indicating a sequential layout). The last three bits of the sub-entry index are used to assign the address translation to its corresponding sub-entry.  In contrast, if there are empty slots among these sub-entries, the pattern is identified as stride. The stride layout is then employed (the layout bit is set to `10'), utilizing the first three bits of the sub-entry index to map the address translation to a sub-entry.
It is important to note that when allocating a new address translation to a sub-entry, it may happen that the sub-entry is already in use by the original base address.  In such cases, we will attempt to relocate the original entry to another sub-entry sharing the same index bits if it is unoccupied. If this alternative sub-entry is also in use, the original address translation that initially occupied the chosen sub-entry will be evicted to accommodate the incoming new address translation. 
Note that the insertion into the L3 TLB is off the critical path and hence does not directly impact performance.

\subsection{Hardware Overhead}

In our configuration, the L3 TLB entries are augmented with additional bits to support the new functionality: a layout bit for sub-entry indexing layout, and an Address Identify Bit (AIB) to identify which address currently occupies the sub-entry. Each TLB entry now comprises two bases, with associated valid/dirty bits, the virtual page base (VPB), AIB, and physical address space (PAS). Therefore, the sharing-aware TLB entry format requires an additional 2 bits (layout bit) + 16 bits (1 bit AIB per sub-entry) + 30 bits (second base address) + 2 bits (v/d for second base address) = 50 bits per TLB entry. Given that our L3 TLB design accommodates 1024 entries, and each entry originally consists of 864 bits (2-bit v/d, 30-bit VPB, and a 52-bit PAS per sub-entry), the introduction of sub-entry sharing and associated metadata increases the size per entry to 914 bits. Additionally, our design adds a 1-bit comparator for each sub-entry to match the AIB. We use CACTI~\cite{cacti} to estimate the area overhead of our approach. The result shows 1.4\% area overhead over the original L3 TLB assuming a 22\,nm technology node.

\section{Evaluation}

\subsection{Overall Performance}
\label{ssec:over_perf}

\begin{figure}[!ht]
\vspace{-6pt}
	\centering
\includegraphics[width=0.48\textwidth]{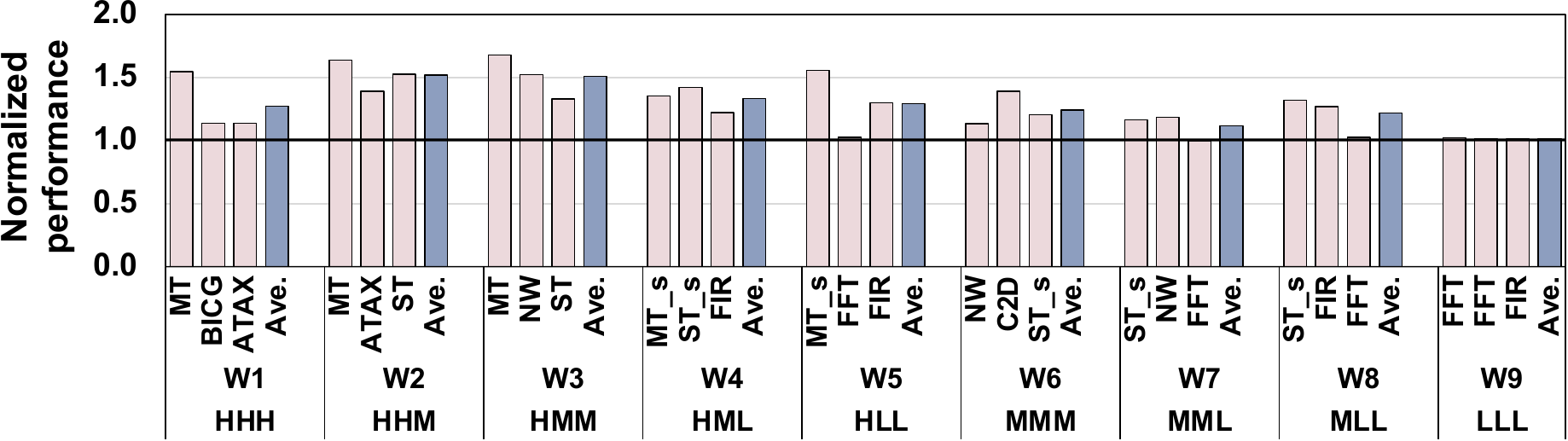}\vspace{-6pt}
\caption{Normalized performance improvements offered by \ourdesign. }
\label{fig:our_perf}
\vspace{-14pt}
\end{figure}

\begin{figure}[!h]
	\centering
\includegraphics[width=0.48\textwidth]{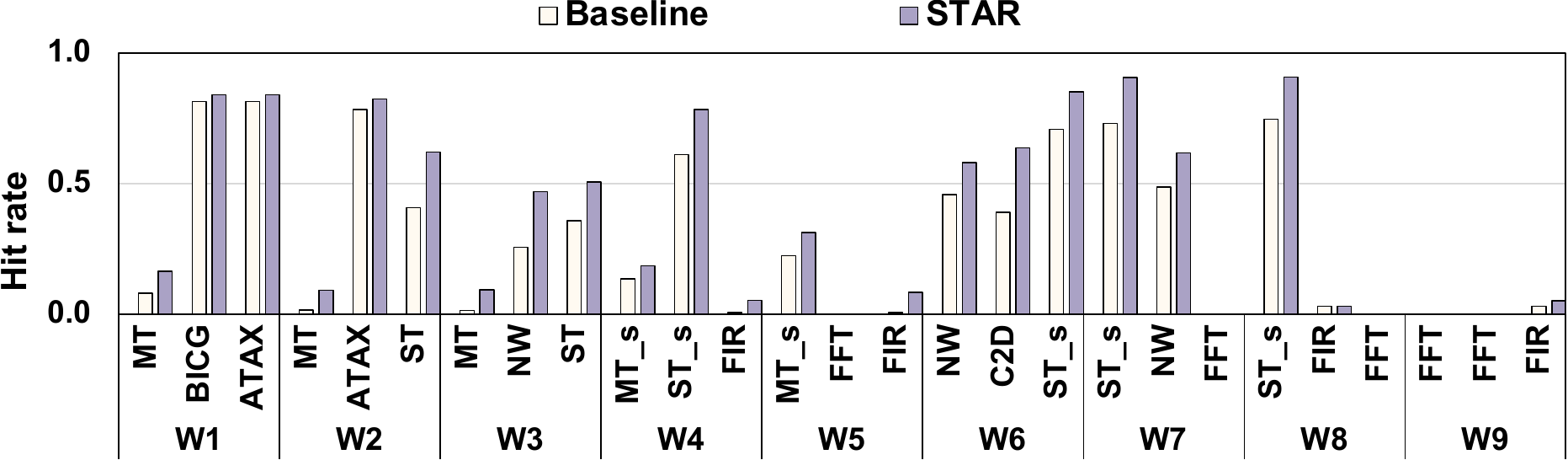}
\vspace{-6pt}
\caption{L3 TLB hit rate of baseline and \ourdesign. }
\vspace{-6pt}
\label{fig:our_hitrate}
\end{figure}

Figure~\ref{fig:our_perf} shows the performance improvements of individual applications within each workload and the harmonic average performance improvements (represented by the last bar of each workload) of the multi-tenant workloads. Results are normalized to the baseline multi-tenant execution.   Figure~\ref{fig:our_hitrate} plots the L3 TLB hit rate for each application in each workload.  One can make the following observations from the figures. First,  the proposed \ourdesign~improves the performance by up to 51.3\%, with an average of 30.2\% across all workloads.
The improvement is more significant for workloads that suffer from severe
contention in the L3 TLB (i.e., high MPKI value). For example, W2 (HHM) achieves 51.3\% improvement and W6 (MMM) achieves 23.5\% improvement, respectively. 
This is because workloads with high MPKI values benefit more from TLB optimizations as each TLB miss leads to costly page table walks, directly impacting performance. Our scheme effectively increases L3 TLB reach by sharing sub-entries, and as a result, the TLB can accommodate more translations and also capture a larger fraction of reused translations.
Interestingly, {\tt FIR} has very low MPKI (0.3) while having 27.1\% performance improvement in W8 (MLL). This is because a large number of pending requests are coalesced to the same TLB miss in L2 MSHR. Reducing handling TLB misses latency can significantly benefit the whole execution.

Second, the performance benefits come mainly from the enhanced L3 TLB hit rates through sub-entry sharing. On average, \ourdesign~improves L3 TLB hit rate by 28.8\% across all workloads. For example, the L3 TLB hit rate of {\tt ST} in W2 improves by 52\%, which translates into a 52.7\% performance improvement. The improved TLB hit rate also indicates an extended TLB reach, effectively reducing the number of (expensive) page table walks.

Third, the performance improvement of the same application is different in different workloads. For example, {\tt MT\_s} achieves a substantial 55\% performance improvement in W5 versus 35\% in W4. This variance can be attributed to how the other applications within these workloads interact with each other. In W5, the stride access patterns of the co-located application {\tt FFT} result in low sub-entry utilization under the baseline scenario. When \ourdesign~is applied, {\tt MT\_s} is able to dynamically share the sub-entries that would otherwise remain underutilized by {\tt FFT}. Since the latter applications do not fully utilize their allocated sub-entries, sharing them with {\tt MT\_s} brings little to no detriment to their performance, hence the significant improvement for {\tt MT\_s}. Conversely, in W4, applications such as {\tt ST\_s} and {\tt FIR} make more efficient use of their allocated sub-entries, leaving fewer opportunities for {\tt MT\_s} to benefit from sharing. 

Finally, our approach does not comprise the performance of any shared applications within the workload. This is because our approach can dynamically shift between shared and non-shared status based on the application demand.  When an application has a higher demand for sub-entries, our approach allows for exclusive access to all sub-entries, the same as the baseline scenario, thus maintaining performance integrity.

\begin{figure}[!t]
	\centering
\includegraphics[width=0.46\textwidth]{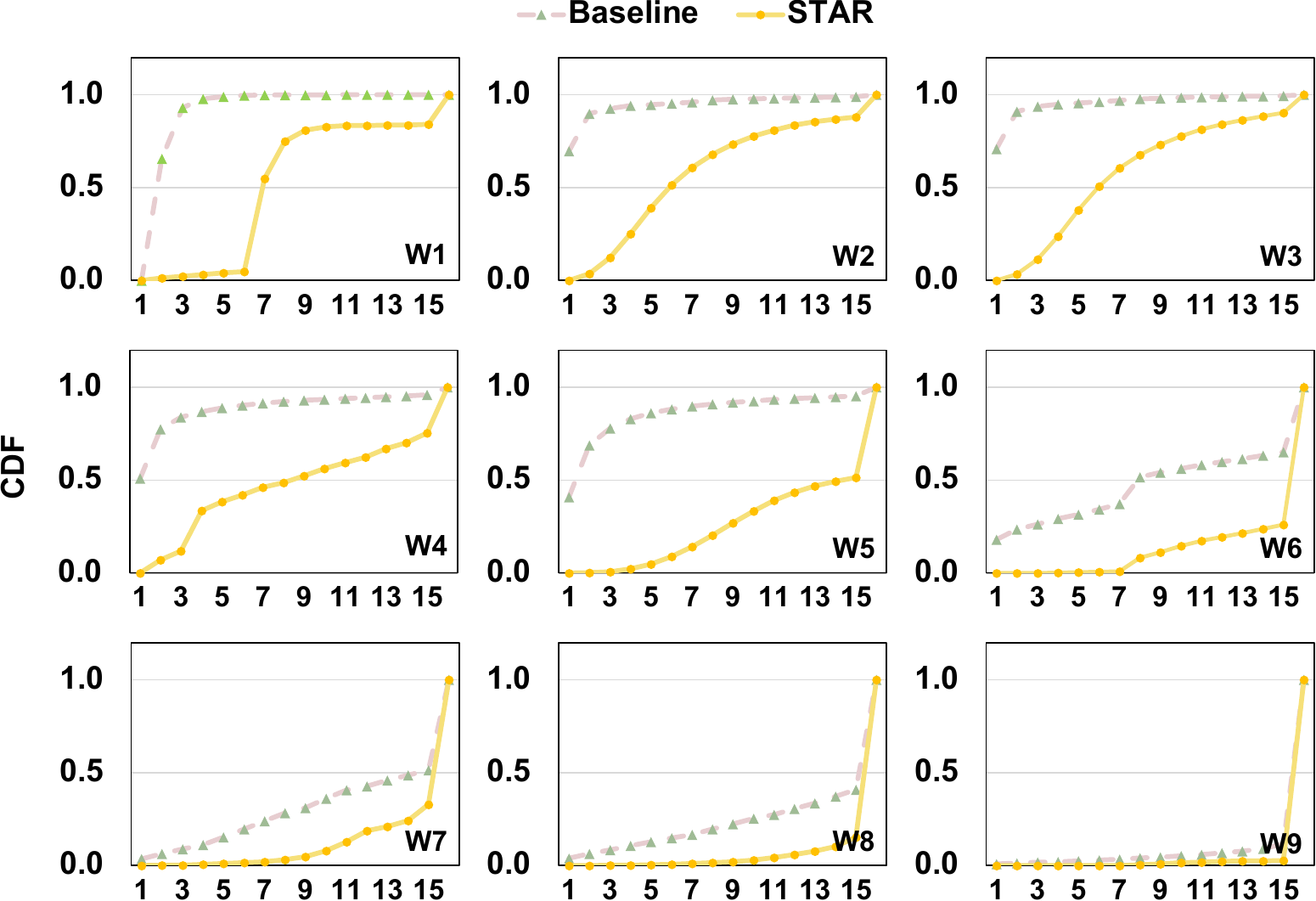}
\vspace{-6pt}
\caption{CDF of sub-entry utilization of \ourdesign.} 
\vspace{-12pt}
\label{fig:our_util}
\end{figure}

We further demonstrate the effectiveness of our approach by plotting sub-entry utilization, as shown in Figure~\ref{fig:our_util}. 
We observe that \ourdesign~consistently achieves higher utilization rates than the baseline, as indicated by the curves of \ourdesign~lying closer to the bottom right compared to the baseline, indicating a larger proportion of TLB entries with higher sub-entry utilization.
We calculate the average utilization by summing up the product of the utilization fraction and the number of occurrences for each eviction and dividing by the total number of evictions. Our approach achieves on average 31.4\% improvement in sub-entry utilization over the baseline. 



\subsection{Sensitivity Analyses}
\label{ssec:sensitivity}

\noindent\textbf{Different number of shared base addresses:}
In our discussion so far, up to two base addresses can share the same TLB entry. We now explore the option of having more base addresses sharing the same entry (i.e., up to 4 base addresses). That is, we allow scenarios where a TLB entry is used by one, two and four base addresses. The 4-base sharing mechanism works as follows. 
For entries with one or two base addresses, the process remains identical to our initial design.
If an entry already has two base addresses and each utilizes fewer than four sub-entries, we enable sharing among four base addresses within that entry.
To facilitate the varied sharing configurations (1, 2 or 4 base addresses), we introduce a 3-bit layout indicator. The initial state `000' indicates no sharing, with the last two bits specifying the sub-entry layout strategy. When the entry is shared, the last two bits deviate from `00' as in our initial design, and the first bit indicates the current number of shared base addresses in the entry. For example, `001' indicates two bases sharing with a sequential layout, while `110' represents four bases sharing with a stride layout.
Our design also incorporates the ability to dynamically transition between non-shared, 2-base address shared, and 4-base address shared states within the TLB. When a base address exhibits an increased demand for sub-entries, and if the entry is currently shared by four bases, we reduce the sharing to two bases, and further one base to accommodate the demand. 

\begin{figure}[!t]
	\centering
\includegraphics[width=0.48\textwidth]{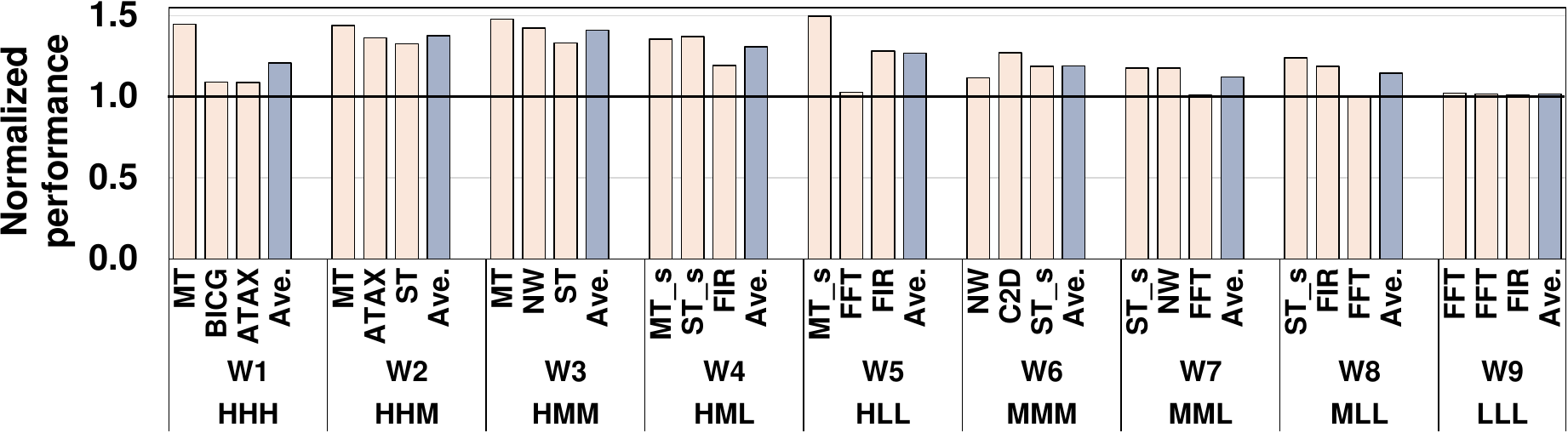}
\vspace{-6pt}
\caption{\ourdesign~with a 4-base sharing TLB. }
\vspace{-12pt}
\label{fig:4_base}
\end{figure}

Figure~\ref{fig:4_base} shows the performance of 4-base sharing normalized to baseline execution. On average, 4-base sharing improves performance by 22.4\% over the baseline. However, it experiences a 7.8\% performance reduction when compared to 2-base sharing.  This is because, 4-base sharing, while enhancing utilization, introduces a trade-off by increasing address conflict evictions. Specifically, four addresses are allocated to share a single sub-entry (compared to two addresses sharing one sub-entry in the initial design), the address conflict evictions increase, which potentially reduces the TLB hit rate. Moreover, the lookup process requires up to four sequential operations in 4-base sharing, further exacerbating the lookup latency.
We also evaluate the hardware overhead of the 4-base sharing approach with CACTI, and it shows a 3.6\% area overhead compared to the baseline.

\begin{figure}[!t]
	\centering
\includegraphics[width=0.48\textwidth]{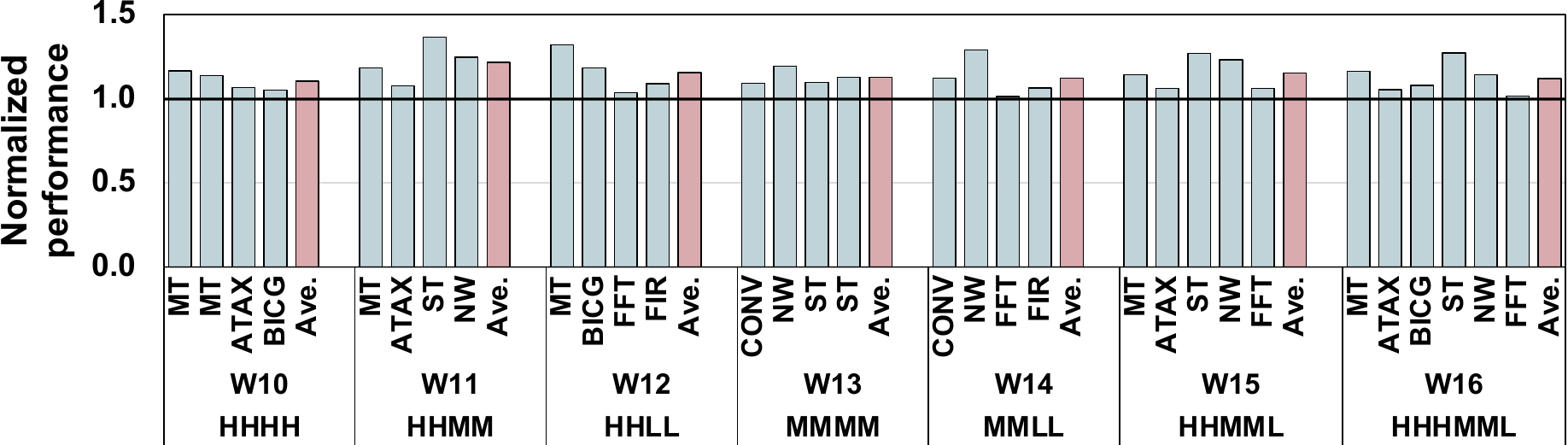}
\vspace{-6pt}
\caption{\ourdesign~with different instance sizes. }
\label{fig:instance_size}
\vspace{-14pt}
\end{figure}

\vspace{1mm}
\noindent\textbf{Different instance sizes:}
We use the applications in Table~\ref{table:app} to form multi-tenant workloads with different numbers of applications, including five workloads with four applications each, one workload with five applications, and one with six applications, as listed in Table~\ref{table:4-instance}. The whole GPU is partitioned into different instance sizes depending on the number of co-running applications, with each instance running one application.  Specifically, in W10-W14 (4-application workload), the GPU is divided into 2+2+2+1; in W15, it is divided into 2+2+1+1+1; and in W16, it is 2+1+1+1+1+1.
Figure~\ref{fig:instance_size} reports normalized performance for \ourdesign. First, our approach is able to deliver scalable performance improvements with different instance sizes, achieving 14.6\%, 15.3\%, and 12.1\% performance improvement in 4-, 5- and 6-application workloads, respectively. Second, the performance improvement is reduced as the number of co-running applications increases. This is because,  first, the decrease in instance size leads to a corresponding reduction in L2 TLB size, which in turn increases the number of requests directed to the L3 TLB. Second, the increase in the number of co-running applications intensifies the competition for the limited number of L3 TLB entries which also impacts performance.


\begin{scriptsize}

\begin{table}[t!]
\scriptsize
  \centering
  
  \caption{Multi-tenancy workloads with 4, 5 and 6 applications.}
\vspace{-2mm}
  \begin{tabular}{|l|l|l|l|}
    \hline
    \textbf{Abbr.} & \textbf{Workload} & \textbf{Applications} & \textbf{Category}\\
    \hline
    \hline
    W10 & workload10 & {\tt MT}, {\tt MT}, {\tt ATAX}, {\tt BICG} & HHHH\\
    \hline
    W11 & workload11 & {\tt MT}, {\tt ATAX}, {\tt ST}, {\tt NW} & HHMM\\
    \hline
    W12 & workload12 & {\tt MT}, {\tt BICG}, {\tt FFT}, {\tt FIR} & HHLL\\
    \hline
    W13 & workload13 & {\tt CONV}, {\tt NW}, {\tt ST}, {\tt ST} & MMMM\\
    \hline
    W14 & workload14 & {\tt CONV}, {\tt NW}, {\tt FFT}, {\tt FIR} & MMLL\\
    \hline
    W15 & workload15 & {\tt MT}, {\tt ATAX}, {\tt ST}, {\tt NW}, {\tt FFT} & HHMML\\
    \hline
    W16 & workload16 & {\tt MT}, {\tt ATAX}, {\tt BICG}, {\tt ST}, {\tt NW}, {\tt FFT} & HHHMML\\
    \hline
    
  \end{tabular}
  \label{table:4-instance}
  \vspace{-6pt}
\end{table}
\end{scriptsize}

\subsection{Comparison to TLB Alternatives}
\label{ssec:other_TLB_designs}

\begin{figure}[!h]
\vspace{-8pt}
	\centering
\includegraphics[width=0.48\textwidth]{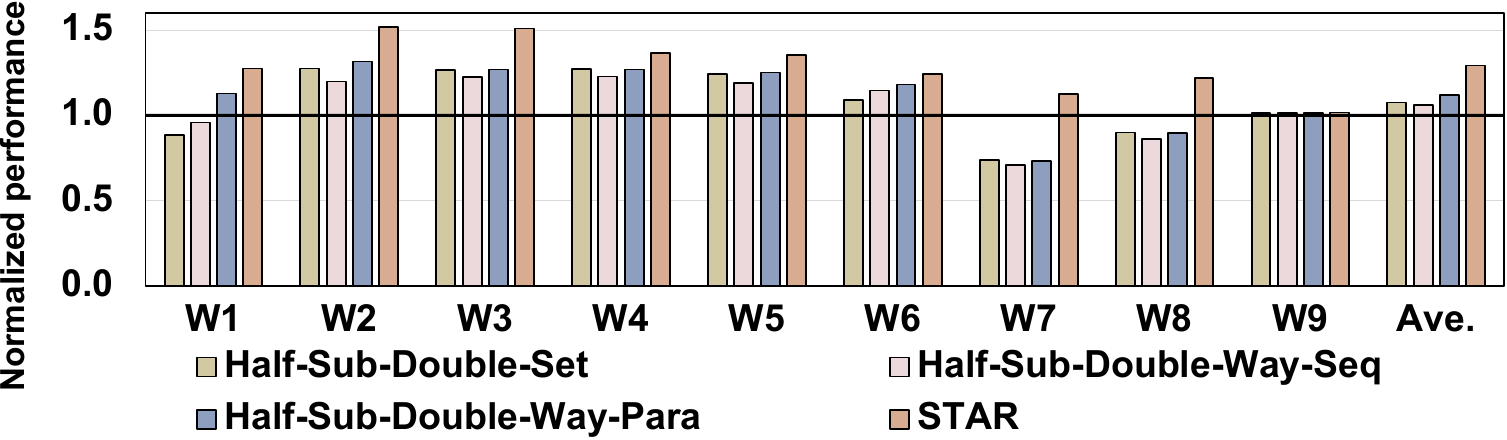}
\vspace{-8pt}
\caption{Comparison of different TLB designs. }
\vspace{-6pt}
\label{fig:TLB_alternative}
\end{figure}


  

    
  

We compare \ourdesign~with three TLB design alternatives, which feature 8 sub-entries per TLB entry while doubling the number of ways or sets to keep total TLB capacity constant relative to the baseline. The TLB entries are exclusively used by one base address. 
Specifically, we consider (i) Half-Sub-Double-Set: 256 sets, 8 ways, and 8 sub-entries per entry; (ii) Half-Sub-Double-Way-Para: 128 sets, 16 ways, and 8 sub-entries per entry --- here we increase the number of comparators with the number of ways such that all ways are compared in parallel within a set; this approach significantly increases hardware overheads; and (iii) Half-Sub-Double-Way-Seq: 128 sets, 16 ways, and 8 sub-entries per entry --- we keep the same number of comparators as the baseline to avoid significant hardware overheads, such that two ways within a set are checked sequentially. We compare the hardware overhead of these alternatives using CACTI: both Half-Sub-Double-Set and Half-Sub-Double-Way-Seq maintain overheads comparable to the baseline with a 1.1\% area increment; while Half-Sub-Double-Way-Para incurs a significant 78.8\% area increment due to the increased number of comparators.

Figure~\ref{fig:TLB_alternative} reports performance for each alternative compared to \ourdesign. All results are normalized to the baseline multi-tenant execution. One can make the following observations. First, \ourdesign~achieves the highest performance improvement among all alternatives. Specifically, our approach achieves a 21.6\%, 23.2\% and 17.4\% performance improvement over Half-Sub-Double-Set, Half-Sub-Double-Way-Seq, and Half-Sub-Double-Way-Para, respectively.
Second, halving the number of sub-entries statically to 8 incurs a performance degradation. For example, in W7, the performance of Half-Sub-Double-Way-Para drops by 26.7\% compared to the baseline. This is because, in the baseline TLB design with 16 sub-entries, a hit in any of the 16 sub-entries reduces the chance of the TLB entry being evicted by the LRU scheme, potentially keeping it in the TLB longer, benefiting other accesses to one of the 16 sub-entries.
This especially benefits the access patterns with good spatial locality, where multiple accesses are closely within a contiguous memory (e.g., {\tt ST}).
In contrast, reducing the number of sub-entries to 8 weakens the capability to exploit spatial locality (i.e., each hit now only benefits 8 sub-entries, compared to 16 in the baseline).
Our approach dynamically alternates between a shared and non-shared TLB status, effectively enhancing the TLB sub-entry utilization while maintaining the efficiency of spatial locality accesses.





\subsection{Comparison to Static TLB Partitioning}
\vspace{-6pt}
\begin{figure}[!h]
	\centering
\includegraphics[width=0.48\textwidth]{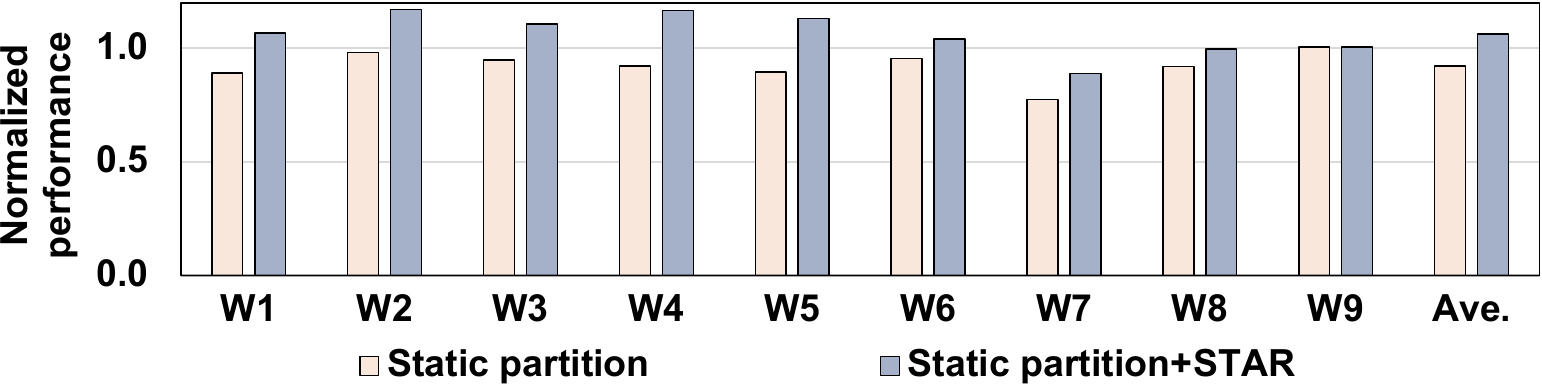}
\vspace{-8pt}
\caption{Comparison to static partition. }
\vspace{-4pt}
\label{fig:partition}
\end{figure}


One straightforward solution to mitigate contention is to statically partition the L3 TLB. In this approach, we statically partition the L3 TLB ways based on instance sizes (i.e., 4-way, 2-way, 2-way for each instance). Figure~\ref{fig:partition} plots the performance of the static partition normalized to the baseline shared L3 TLB. We observe an average 7.9\% performance degradation due to static partitioning.  Workloads combining applications with mixed MPKI values, particularly those including at least one high/medium-MPKI application (such as in W5, W7, and W8) suffer from a more severe performance drop. This is because static partitioning restricts the number of TLB entries available to high/medium-MPKI applications, thus reducing the ability of applications to accommodate increasing demands by taking up entries from others.
\ourdesign~is also adaptable to scenarios with static partitioning, enabling two base addresses within the same instance or process to share a single TLB entry.  Figure~\ref{fig:partition} also shows the performance of our approach on top of the static partitioning. The results are normalized to the baseline TLB sharing execution. \ourdesign+static partitioning achieves an average of 14.2\% performance improvement over static partitioning alone. This is because our approach is able to further optimize sub-entry utilization within individual processes, effectively increasing the TLB hit rate and enhancing overall performance.

\subsection{Comparison with State-of-the-Art}

\begin{figure}[!h]
	\centering
\includegraphics[width=0.44\textwidth]{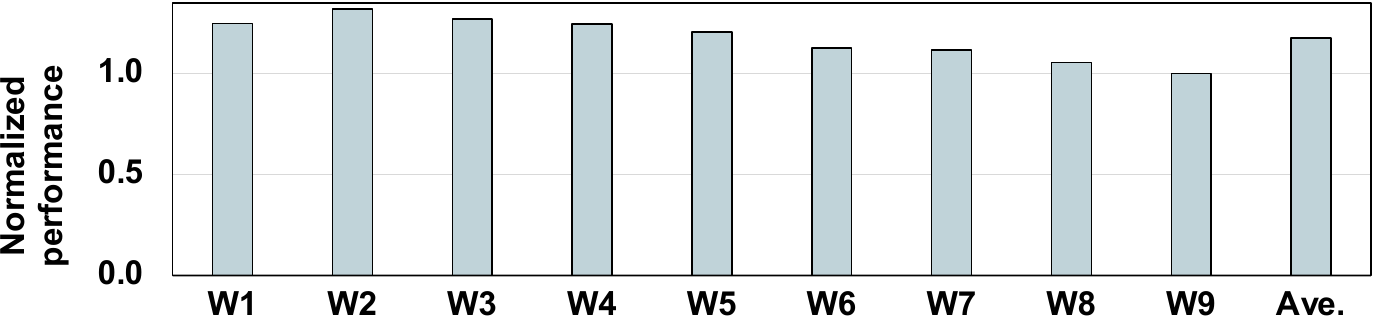}
\vspace{-8pt}
\caption{Combined with MASK~\cite{Ausavarungnirun:2018:MRG:3173162.3173169}. }
\vspace{-16pt}
\label{fig:mask}
\end{figure}

The previous work MASK~\cite{Ausavarungnirun:2018:MRG:3173162.3173169} addressed shared TLB contention in multi-application environments using TLB-Fill Tokens to manage how many warps can fill the shared TLB, and adjusted the TLB entries allocated to each application based on its L2 TLB miss rate, thus reducing thrashing. It also features a TLB bypass cache for entries from warps with insufficient tokens. 
Although MASK effectively manages TLB contention through dynamic partitioning, it lacks optimizations at the sub-entry level. This suggests that \ourdesign~is complementary to MASK. Figure~\ref{fig:mask} shows the performance improvement of MASK+\ourdesign~normalized to MASK. We note an average 17.6\% performance improvement over MASK. This demonstrates that \ourdesign~can work with TLB dynamic partitioning optimizations, bringing additional performance benefits.
\section{Related Work}
Substantial prior studies have focused on address translation optimizations to improve system performance~\cite{li2021improving, Alam:2017:DVM:3079856.3080209, Barr:2010:TCS:1815961.1815970,6307767,Bhattacharjee:2017:TP:3037697.3037705, Kumar:2018:LLT:3173162.3173198,8192492,6835964}.
Several previous studies~\cite{6307767,pham2015large} enhanced TLB hit rates by employing speculative techniques to predict the translations that miss in the TLBs.
Many studies have delved into methods designed for enhancing page management to optimize the address translation process~\cite{7056046,agarwal2015page,marathe2006hardware,yan2019nimble,agarwal2017thermostat, li2023idyll, wang2024grit, HPCAli}.
An alternate set of techniques~\cite{Basu:2013:EVM:2485922.2485943, gandhi2014efficient, Karakostas:2015:RMM:2749469.2749471} improved the TLB reach by generating contiguous translations. 
Research proposals also suggested an alternative memory management unit (MMU) cache structures, to cache multiple levels of the page tables~\cite{Barr:2010:TCS:1815961.1815970, bhattacharjee2013large}.
Additionally, the research community has explored approaches to increase TLB performance by using large pages and improving super-page management~\cite{du2015supporting, panwar2018making}. 
Bharadwaj et al.~\cite{8574547} co-designed distributed TLBs with a lightweight interconnect to realize scalable shared L2 TLBs.
Li et al.~\cite{li2021improving} optimized address translation in multi-GPUs through sharing and spilling aware TLB design. 
Achermann et al.~\cite{achermann2020mitosis} proposed Mitosis, a hardware optimization to reduce address translation overheads by eagerly keeping page tables local through replication and migration.
Compared to all the prior efforts, our research pioneers the optimization of MIG-enabled GPUs by innovatively addressing TLB thrashing and implementing a sharing mechanism for advanced TLB sub-entry designs.

\section{Conclusion}

In this paper targeting multi-instance GPUs, we comprehensively study the address translation efficiency in multi-tenant execution.  Our investigation reveals that shared L3 TLB contention significantly impacts performance by increasing TLB thrashing and reducing the utilization of TLB sub-entries. To address this problem, we propose \ourdesign~that enables dynamic sharing of TLB entries among different base addresses.  Experimental results demonstrate that \ourdesign~substantially enhances performance, delivering an average improvement of 30.2\% across a variety of multi-tenant workloads.


\balance
\bibliographystyle{IEEEtranS}
\bibliography{refs}

\end{document}